\documentclass{JHEP3}

\skip\footins = 1\bigskipamount plus 2pt minus 4pt

\usepackage{epsfig}
\usepackage{graphicx}

\newcommand{\IR}{\mathbb{R}}
\newcommand{\IS}{{\bf S}}

\def\half{\frac{1}{2}}
\def\del{{\partial}}
\def\room{~\rule[-2mm]{0mm}{8mm}}

\def\Schw{Schwarzschild}
\def\Lp{L_{\rm poles}}
\def\({\left(}
\def\){\right)}
\newcommand\SO{\mathop{\rm SO}\nolimits}
\preprint{{\tt hep-th/0406002}}
\title{A dialogue of multipoles: matched asymptotic expansion for caged black holes}

\author{Dan Gorbonos and  Barak Kol\\
Racah Institute of Physics, Hebrew University\\
Jerusalem 91904, Israel\\
E-mail: \email{gdan@phys.huji.ac.il},
{\tt\href{mailto:barak_kol@phys.huji.ac.il}{barak\_kol@phys.huji.ac.il}}}

\abstract{No analytic solution is known to date for a black hole in a
  compact dimension. We develop an analytic perturbation theory where
  the small parameter is the size of the black hole relative to the
  size of the compact dimension. We set up a general procedure for an
  arbitrary order in the perturbation series based on an asymptotic
  matched expansion between two coordinate patches: the near horizon
  zone and the asymptotic zone. The procedure is ordinary perturbation
  expansion in each zone, where additionally some boundary data comes
  from the other zone, and so the procedure alternates between the
  zones. It can be viewed as a dialogue of multipoles where the black
  hole changes its shape (mass multipoles) in response to the field
  (multipoles) created by its periodic ``mirrors'', and that in turn
  changes its field and so on. We present the leading correction to
  the full metric including the first correction to the
  area-temperature relation, the leading term for black hole
  eccentricity and the ``Archimedes effect''.  The next order
  corrections will appear in a sequel. On the way we determine
  independently the static perturbations of the Schwarzschild black
  hole in dimension $d \ge 5$, where the system of equations can be
  reduced to ``a master equation'' --- a single ordinary differential
  equation. The solutions are hypergeometric functions which in some
  cases reduce to polynomials. }

\begin{document}

\section{Introduction and summary}

\subsection{Background}

In the presence of extra compact dimensions, there exist several
phases of black objects, namely massive solutions of General
Relativity, depending on the relative size of the object and the
relevant length scales in the compact dimensions. For concreteness, we
consider a background with a single compact dimension --- $\IR^{d-2,1}
\times \IS^1$, where $d$ is the total spacetime dimension and $d \ge
5$ (in order to avoid spacetimes with 2 or less extended spatial
dimensions where the presence of a massive source is inconsistent with
asymptotic flatness).

In this background one expects at least two phases of black object
solutions: when the size of the black object is small (compared to the
size of the extra dimension) one expects the region near the object to
closely resemble a $d$-dimensional black hole, while as one increases
the mass one expects that at some point the black hole will no longer
fit in the compact dimension and a black string, whose horizon winds
around the compact dimension will be formed. Thus we distinguish
between the black hole and the black string according to their horizon
topology which is either spherical --- $\IS^{d-2}$ or cylindrical ---
$\IS^{d-3} \times \IS^1$, respectively. We shall sometimes refer to
such a black hole in a compact dimension as a ``caged black hole''.

More generally one could consider backgrounds $\IR^{d-c-1,1} \times
X^c$ where $X^c$ is any $c$-dimensional compact Ricci-flat manifold
such as the $c$-dimensional torus, the K3 surface or a Calabi-Yau
3-fold. For a general $X$ more black object phases will exist, but we
expect that generically the essential phase transition physics between
any two specific phases will be qualitatively similar to the $X=\IS^1$
case.

This system raises several deep questions in general
relativity~\cite{TopChange}: during the real-time phase transition a
naked singularity may be encountered which may require intervention
from quantum gravity and an amendment to Cosmic Censorship; the
transition would certainly produce some sort of a
``thunderbolt''~\cite{HawkingStewart,explosive}; the system exhibits a
critical dimension for stability in at least two
instances~\cite{TopChange,SorkinCriticalDim}; it is a prototype
example for the failure of black hole uniqueness in higher dimensions
$d \ge 5$~\cite{BlackRing,uniqueness}; a novel kind of topology change
is expected to play a central role~\cite{TopChange}; and there is an
ongoing debate regarding the correct phase diagram, especially whether
a stable non-uniform\footnote{not invariant under translation along
  the compact direction.}  string phase exists (for all
$d$)~\cite{HorowitzMaeda1}.

Considerable work, much of it recent, went into finding the various
solutions in this background. A uniform black string is readily
described analytically by an arbitrary mass \Schw\ solution in $d-1$
dimensions with an added spectating coordinate. Gregory and Laflamme
(GL,1994) discovered that this solution develops a tachyon below a
certain critical mass~\cite{GL1}. Horowitz and Maeda (2000) gave an
argument for the existence of stable non-uniform
strings~\cite{HorowitzMaeda1}. Gubser analytically perturbed the
critical GL string to find approximate expressions for non-uniform
solutions (which we interpret to be unstable)~\cite{Gubser}. De-Smet
attempted to find analytic solutions by classifying 5d algebraically
special metrics (the same method Kerr used successfully in 4d to
obtain the rotating black hole) not finding any novel ones in this
background~\cite{deSmet}. Harmark and Obers found a clever ansatz that
reduces the number of unknown metric functions from 3 to 2, but could
not completely solve the equations either~\cite{HO1}.  Recently Sorkin
generalized the analysis of Gubser from 5d (and 6d which was done by
Wiseman~\cite{Wiseman1}, fixing some problems with the earlier
analysis) to an arbitrary dimension, discovering an interesting
critical dimension: for $d \ge 14$ the non-uniform branch emanating
from the GL point changes its nature and essentially becomes
stable~\cite{SorkinCriticalDim}. A critical dimension for a different
point in the phase diagram was already predicted in~\cite{TopChange}.

The limited success of analytical methods created a demand for
numerical solutions. The branch of non-uniform black string solutions
emanating from the GL point was obtained numerically by
Wiseman~\cite{Wiseman1} (see also earlier work in~\cite{WisemanStars}
and a post-analysis in~\cite{Wiseman2,KolWiseman}) who managed to
formulate axially-symmetric gravitostatics (namely, essentially 2d) in
a ``relaxation'' form (a procedure familiar from electrostatics) while
presenting the constraints through ``Cauchy-Riemann --- like"
relations. Even though there is no definitive answer yet whether these
solutions are indeed unstable, the author argued that irrespectively
they cannot serve as an endpoint for the decay of the GL string. While
all the solutions we mentioned above are static Choptuik et
al. performed a demanding numerical time evolution for the decay of
the black string, but had to stop before the end state was reached due
to an essential limitation of the algorithm used (grid stretching) in
the high curvature region which forms~\cite{CLOPPV}.

More recently the focus shifted from black strings to black holes.
While intuition leads us to expect that a small black hole should
exist being indifferent to the existence of a much larger compact
dimension, no analytic solution is available to
date. In~\cite{PiranSorkin} indications for the nature of the phase
transition were gained from an analysis of possible time-symmetric
initial data. In~\cite{KudohTanakaNakamura} the closely related
problem of black holes in a braneworld was tackled
numerically. In~\cite{HO2,KPS1} it was shown that indeed there are
order parameters such that the black hole and black string are at
finite values, as was assumed in~\cite{TopChange}, and
moreover~\cite{KPS1} announced most of the quantitative results of the
current paper. In~\cite{KPS2,KudohWiseman} numerical black hole
solutions were presented for the first time in 5d and 6d respectively,
giving strong evidence for their existence. Finally,~\cite{H4}
presented a ``first order analytic approximation'' of small black
holes in the framework of the Harmark-Obers coordinates~\cite{HO1},
and~\cite{Karasik:2003tx} found an analytic approximation for a small
black hole on a brane.

\subsection{Motivation and basic set-up}

In this paper we present the first analytic (though perturbative)
procedure to obtain solutions for small black holes (BH's). Let us
introduce some notation (see figure~\ref{coordinates}). We denote by
$z,\, r$ the ``cylindrical'' coordinates, where $z$ is the coordinate
along the compact dimension whose period we denote by $L$, and $r$ is
the radial coordinate in the extended $\IR^{d-2}$ spatial
dimensions. The problem is characterized by a single dimensionless
parameter, for instance the dimensionless mass $\mu \equiv G_N
M/L^{d-3}$ where $G_N$ is the $d$-dimensional Newton constant and $M$
is the mass (measured at infinity), or $\mu_\beta\equiv\beta/L$ where
$\beta$ is the inverse temperature.  In the vicinity of the black hole
it is useful to introduce ``spherical'' coordinates $\rho,\, \chi$ as
well. We denote by $\rho_0$ the \Schw\ radius of the BH (in the small
BH limit), and one has $\rho_0^{~d-3} = {\rm const}\, G_N \, M$ where
${\rm const}$ is a dimensionless constant.

{\renewcommand\belowcaptionskip{-1em}
\FIGURE[t]{\centerline{\epsfig{file=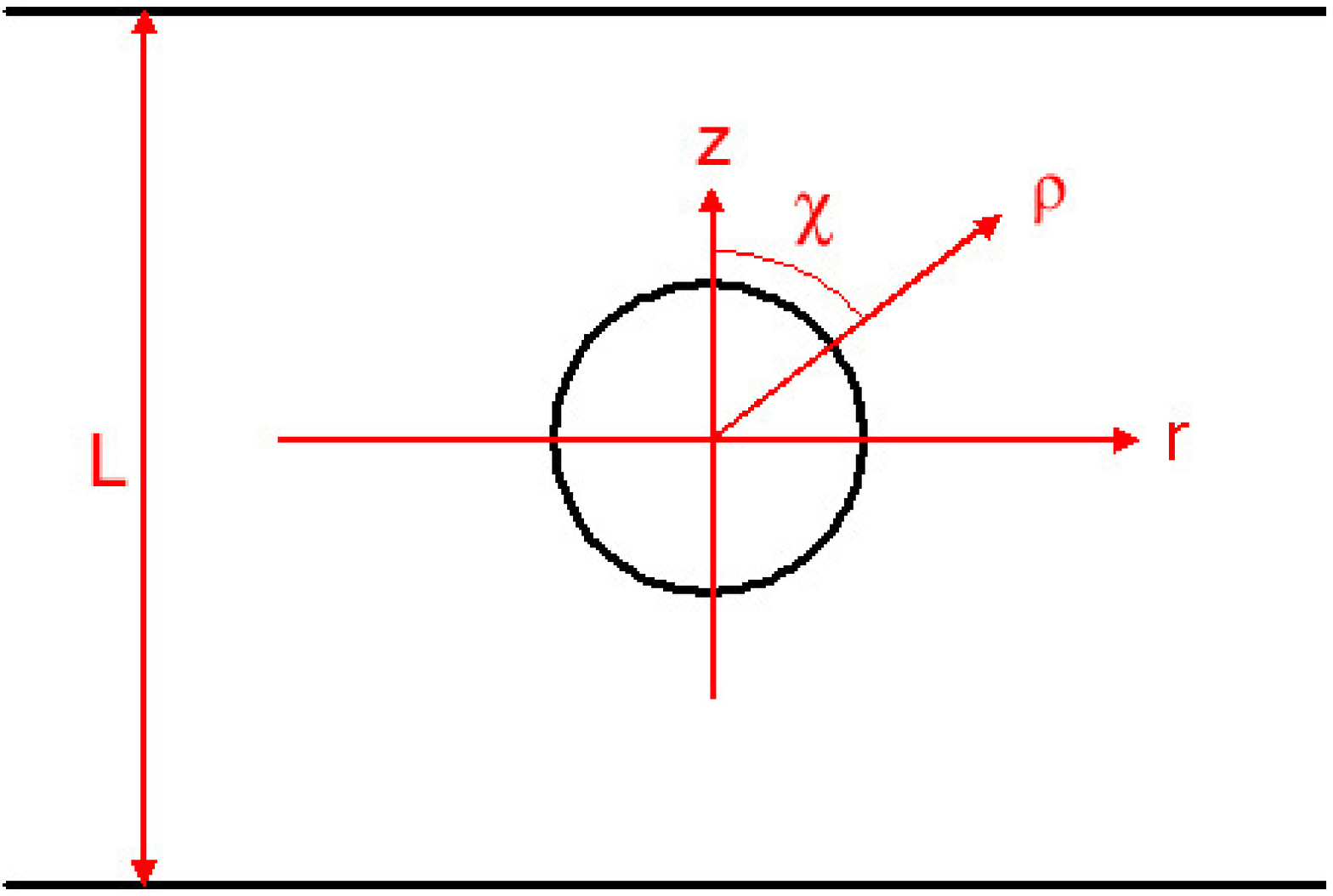, width=.48\textwidth, clip=}}%
\caption{Illustration of the $(r,z)$ ''cylindrical" coordinates and
  the $(\rho,\chi)$ ''spherical" near horizon coordinates. The period
  of the compact dimension (in the $z$ direction) is denoted by
  $L$.\label{coordinates}}}
}

There is reason to expect good analytic control of small black holes
even if we do not have a complete analytic solution since we have two
good approximation in two different regions which overlap: for $\rho
\ll L$ the metric is expected to resemble closely a $d$-dimensional
\Schw-Tangherlini BH~\cite{tangher}, while for $\rho \gg \rho_0$ the
gravitational field is weak and the newtonian approximation
holds. Hence $\mu$ or more precisely $\rho_0/L \propto \mu^{1/d-3}$ is
our small parameter for the perturbation.

The motivations for this research are first to obtain a theoretical
description of this simple system which is important on its own right,
and second to gain understanding of the phase transition physics
through combination with numerical work. The symbiosis with numerical
work comes close to serve as a partial substitute of experiments
(which are sorely absent in this field): the numerics are essential
for understanding big black holes close to the phase transition where
the perturbative expansion is expected to break down, and the analytic
control serves to formulate the aims and methods of the
numerics. Moreover, the two can be used to test and confirm one
another, as was the case for this research and~\cite{KPS1,KPS2}. As it
turns out the largest BH's obtained numerically show only a single
multipole mode correction to their spherical
horizon~\cite{KPS2,KudohWiseman}, and that lends some hope that the
analytic expansion would retain some validity for large BH's as well.

\paragraph{Basic set-up.}

The first decision to be made it to choose the coordinates and ansatz
for the analysis. At first one would hope to use a single coordinate
patch for the whole metric. However,~\cite{KudohTanakaNakamura} showed
that in the popular conformal coordinates (where the metric in the
$(r,z)$ plane is in conformal form $ds^2 = e^{2 \hat{B}(r,z)}\,
\left(dr^2+dz^2\right)$, see also below~(\ref{defx})) the coordinate
size of the horizon is a conformal invariant and hence the coordinate
patch necessarily changes with $\mu$. A similar phenomenon happens in
the Harmark-Obers coordinates which are a semi-infinite cylinder with
a single marked point [$(r,z)=(0,L/2)$] where the coordinate
transformation is singular and whose location changes with $\mu$.

Therefore we choose to work with \emph{two coordinate
  patches} (see figure~\ref{zones}): the \emph{near zone} $\rho \ll L$
where the horizon ($\rho_0$) is fixed and the periodicity of $z$ is
invisible far away, and the \emph{asymptotic zone} $\rho \gg \rho_{0}$
where $L$ is fixed and $\rho_0$ is invisible. The metric in the two
regions must be consistent over the overlap region $\rho_0 \ll \rho
\ll L$ (which grows indefinitely as $\mu \to 0$).

\pagebreak[3]

\FIGURE[t]{\epsfig{file=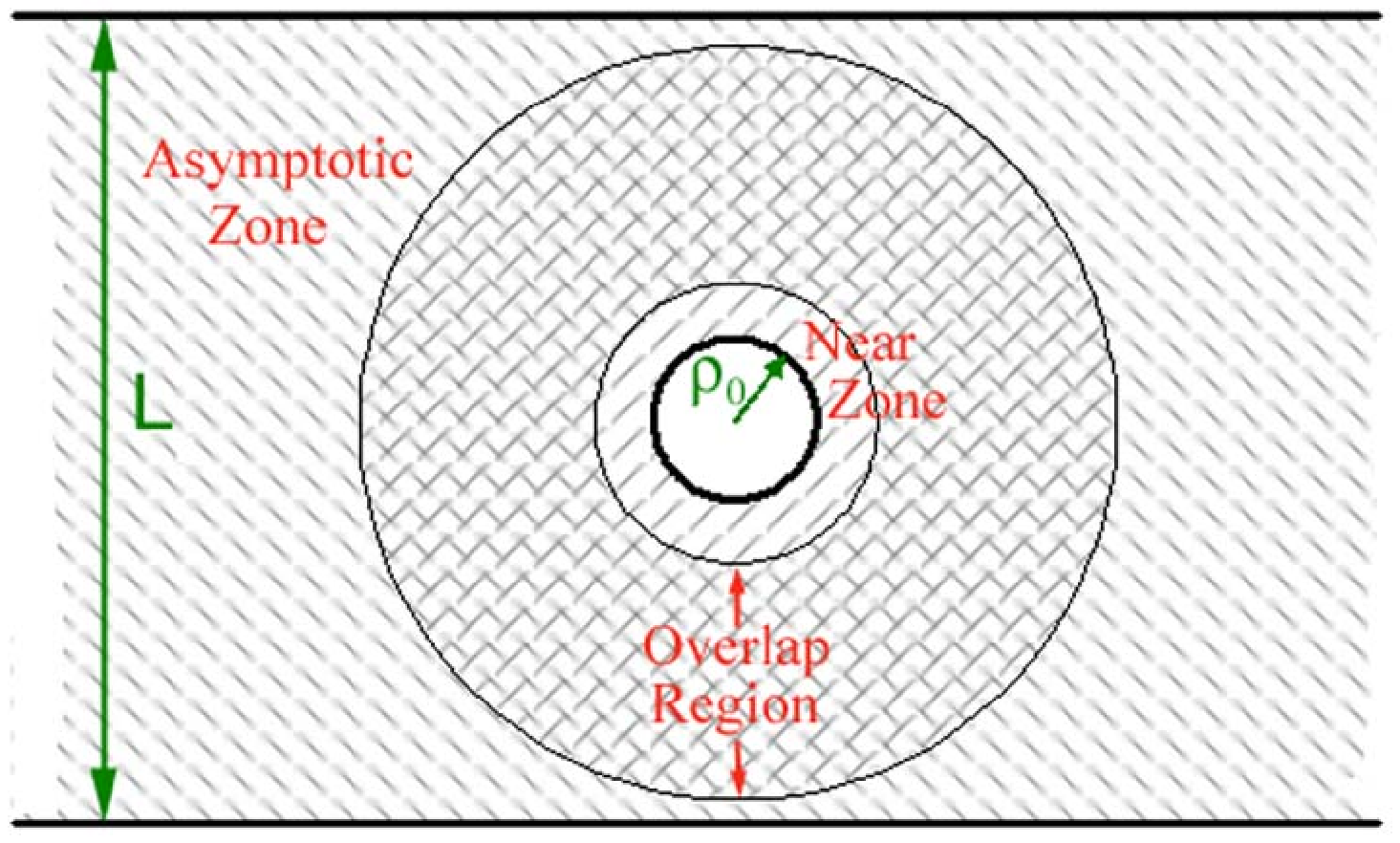, width=.6\textwidth, clip=}%
\caption{The division of the spacetime into two zones: the
  \emph{near zone} $\rho \ll L$ where $\rho_0$ is fixed and the
  perturbative parameter is $L^{-1}$, and the \emph{asymptotic zone}
  $\rho \gg \rho_{0}$ where $L$ is fixed and the perturbative
  parameter is $\rho_{0}$. The two zones overlap over the
  \emph{overlap region} which increases indefinitely in the small
  black hole limit. During the perturbation process the two zones are
  separate, and communicate only through the matching
  ``dialogue''. The near zone is defined by $\{ (\rho,\chi): \, \rho
  \ge \rho_0 \}$ while the asymptotic zone is defined by $\{ (r,z):\, r
  \ge 0, ~z \sim z + L \} \backslash (0,0)$.\label{zones}}}

Such a procedure is known in General Relativity as a {\it ``matched
  asymptotic expansion''} --- the metric is solved for in each
asymptotic region and certain quantities are determined by matching
the metrics over the overlap (for some recent examples
see~\cite{Alvi,Poisson,Poujade}, in mathematical physics this idea
goes back as far as Laplace who used it to find the shape of a drop of
liquid on a surface --- see~\cite{Damour} and references therein for a
historical review). However, this is probably the first time such a
procedure is used to find a static black hole solution.

We start by defining the domain for each zone and the zeroth order
solution. In the near zone, whose domain is $\{ (\rho,\chi): \,\rho
\ge \rho_0 \}$ we have a $d$-dimensional \Schw\ black hole metric,
with fixed $\rho_0$ and the perturbation is in orders of $L^{-1}$. In
the asymptotic zone, on the other hand, the domain is $\{ (r,z):\, r
\ge 0,\, z \sim z + L \} \backslash (0,0)$, the zeroth order solution
is simply the flat ``cylinder'' with the origin omitted, and the
perturbation parameter is $\rho_0$.

Our objective is to describe the perturbation process (to any order)
and apply it. The paper is organized as follows. In
section~\ref{newtoniansec} we describe the equations in the asymptotic
zone and especially the newtonian potential. In
section~\ref{BHperturb} we determine the linearized corrections to the
\Schw\ black hole in the near zone. In
section~\ref{MatchingProcedure-section} we describe the general
perturbation procedure for this system and in~\ref{results} we present
quantitative results on the leading corrections to the zeroth order
metric. In the appendices we review some information on Heun's
equation and the Hypergeometric equation, review the definition of the
surface gravity and give some details on vector harmonics in 5d (on
$\IS^3$). We now turn to the summary.

\subsection{Summary}
\paragraph{The asymptotic zone.}

In section~\ref{newtoniansec} we describe the asymptotic zone. The
first correction to the zeroth order metrics described above is
readily computed --- it is the newtonian approximation in the
asymptotic region (in standard harmonic gauge), where the newtonian
potential~(\ref{Newtonian-potential}) is obtained through the method
of mirror images (considering the infinite sequence of sources in the
covering space at $z=n L$ for any integer $n$). This term is
proportional to $G_N\, M \propto \rho_0^{~d-3}$ and hence belongs to
order $d-3$ in the asymptotic zone. Actually, the whole post-newtonian
procedure is relevant and we review it, though in this paper all we
need is the lowest order (newtonian) approximation.

\paragraph{Black hole perturbations.}

The next correction to consider is the leading correction to the
\Schw\ solution. It is the analogue of the newtonian approximation
only here the unperturbed background is curved, and it describes the
response of the geometry to the mirror sources far away.  Despite the
analogy with the newtonian approximation the implementation is
involved and is described in section~\ref{BHperturb}.

In 4d this computation was carried out by Regge and
Wheeler~\cite{Regge} who solved for all the linear perturbations, not
only the static ones. Recently Ishibashi and Kodama succeeded to
generalize their result to an arbitrary dimension, and to include a
cosmological constant as well~\cite{kodama2,kodama3,kodama4}.  Our
treatment is independent and we compare the two approaches below after
describing our own.

The computation involves a few steps. We start by writing down the
most general static perturbation to the metric. The spherical symmetry
guarantees that at linear order perturbations in different
representations of the rotation group will not mix, and we find by
counting degrees of freedom that it suffices to consider ``scalar
harmonics'' --- representations which are the symmetric product of the
vector representation, or equivalently, metric perturbations which are
determined by scalar functions on the sphere. It turns out that the
spherical symmetry also suggests a natural gauge which we term
{\it``no derivatives gauge''} and completely fixes the
reparameterization invariance, leaving us with 3 undetermined metric
functions (fields).

Writing down the equations of motion and separating the angular
variables we find that a Ricci flatness condition in the angular
directions yields an algebraic relation among the radial
functions~(\ref{algeb}) which is similar to a trace condition and
allows us to eliminate one of the fields. After substitution one can
express one of the remaining fields in terms of the other and its
first and second derivatives. Performing the second substitution we
are left with a second order ordinary differential equation (ODE),
rather than a third or fourth order one would initially expect. So
finally one has a single second order ODE in the radial direction, for
one metric function (and for each spherical harmonic mode) from which
the whole metric may be recovered. This is the so called
\emph{``master equation''} which after a change of variables
simplifies further to become~(\ref{master-eq}). It would be nice to
have a deeper understanding why these reductions were to be expected.

The master equation belongs to the \emph{Heun class of Fuchsian
  equations}, where Fuchsian means that the equation has only
regular-singular points on the complex sphere which includes infinity,
and Heun means that there are exactly 4 such points.  Unlike the
Hypergeometric case of 3 regular singularities there is no general
solution to the Heun equation, though several methods are
available. In this case however, it turns out that the solutions can
be written in terms of \emph{a hypergeometric function}, and it would
be nice to understand why that had to be the case. Interestingly, we
observe that in some of the relevant cases these hypergeometric
functions simplify further to {\it polynomials}, and in particular in
5d all relevant solutions are polynomials (solutions which are of even
multipole number and are regular at the horizon).

We started working out this problem before we were aware of the
results of Kodama and Ishibashi~\cite{kodama2,kodama3,kodama4} and we
continued independently even after learning about these papers in
order to avoid the formalism of gauge invariant perturbation theory,
and the various changes of variables which are employed there. We were
able to do so and actually found a somewhat different master (Heun)
equation. Yet the final reduction of our master Heun equation to a
hypergeometric one was motivated by those papers.

\paragraph{The matching procedure.}

\looseness=1 One of the main results of this paper is the construction of a
perturbation method for the metric (in both patches) which may be
carried in principle to an arbitrarily high order in the small
parameter. The method is described in
section~\ref{MatchingProcedure-section}. A crucial step is to identify
\emph{a dimensionful expansion parameter on each patch}: $\rho_0$ in
the asymptotic zone and $L^{-1}$ in the near region. As in any
perturbative expansion, at each order one needs to solve a
non-homogenous linear equation --- the linear equation being the same
as the one which appears at first order and the non-homogeneous source
term being constructed from lower order metric functions. The precise
form of the source term depends on the higher order gauge choice which
we do not specify, but will not change the method we describe. The
solution to this equation is determined up to a solution of the
homogeneous equation. This indeterminacy for each zone on its own
reflects the freedom of adding external field multipoles --- in the
asymptotic zone they are situated at the origin while in the near zone
they are at infinity. These external multipoles must be determined by
matching with the other zone, a procedure which requires to identify
(after matching the gauge) the leading terms in the metric on both
zones.  We call this process \emph{``a dialogue of multipoles''}.

A priori it is not obvious that the required terms from the other zone
are already available at the right time, namely that \emph{the method
  is well-posed} (that there are sufficient boundary
conditions). Hence it is interesting to study at any given order in a
specific zone which orders must be already available for matching from
the other one, and thereby describe the pattern of the dialogue ---
the orders at which one should alternate between the zones.  This
pattern can be determined by a simple dimensional analysis of the
multipole coefficients as we describe in the text, and indeed we find
the system to be well-posed. Interestingly, \emph{the dialogue pattern
  which emanates is dimension dependent}: 5d is special in that one
scales a single order in the perturbation ladder on each zone and then
alternates to the other zone; for $d>5$ one needs to climb several
steps before going to the other zone, and in the $d \to \infty$ limit
one gets infinitely many constants already from matching with the
newtonian potential alone.

\paragraph{Quantitative matching results.}

In section~\ref{results} we apply the general procedure to the leading
order and match the newtonian potential at the asymptotic zone to get
the leading correction to the \Schw\ solution in the near zone. To
this purpose it is essential to have available certain matching
constants which can be read from the explicit solutions we derived for
the linear perturbations of Schwarzschild. The next order correction
is currently under study~\cite{GKin-progress}.

>From the metric which we obtain one may extract certain
``measurables'':
\begin{itemize}
\item The leading correction to the mass --- temperature relation is
  given in~(\ref{Akappa}). At this order the BH is still spherical but
  there is a correction to this relation since the small black hole
  does not asymptote in the near zone to flat space with zero
  potential, but rather there is a non-zero potential shift due to the
  images.
\item The leading (quadrupole) departure from a spherical horizon ---
  measured by the ``eccentricity'' is given
  in~(\ref{eccentricity-rho0}). The result of the deformation is to
  make the black hole longer along the $z$ axis compared to the $r$
  axis as in figure~\ref{eccentricity} and can be understood from the
  shape of small (newtonian) equipotential lines around $\rho=0$ (see
  figure~\ref{black2}).
\item The coefficient of the ``inter-polar distance'' is given
  in~(\ref{defId}).  By ``inter-polar'' distance we mean the proper
  distance from the ``north pole'' of the black hole around the
  compact circle and up to the ``south pole''(see
  figure~\ref{lpoles}). Actually the black hole tends to ``make room''
  for itself, in the sense that the inter-polar distance added to the
  black hole size in conformal coordinates is always larger than $L$,
  the size of the compact dimension. This can be re-stated as the
  observation that such black holes seem to always have a positive
  scalar charge as seen from infinity similar to the ordinary positive
  mass theorem (where the scalar is the one which arises from the
  dimensional reduction of the $g_{zz}$ metric component --- the size
  of the extra dimension). In 5d the effect is the strongest, where to
  leading order in the small parameter the inter-polar distance does
  not decrease at all. We term that ``a black hole Archimedes effect''
  since the black hole repels or expands an amount of space equal to
  its size in 5d (and less in higher dimensions).
\end{itemize}

Most of these results were already announced in~\cite{KPS1} and here
we add the determination of the inter-polar distance and the
generalization of the eccentricity for $d>5$.  They were numerically
confirmed in 5d~\cite{KPS2} as well as in
6d~\cite{KWprivate,Sorkin-private} and other
dimensions~\cite{Sorkin-private}. Recently a paper~\cite{H4} has
appeared deriving the leading order form of the metric within the
framework of the Harmark-Obers coordinates~\cite{HO1}, and as such
overlaps with the results announced in~\cite{KPS1} and proven here.
The overlap includes the corrections to the temperature and area,
while~\cite{H4} obtains also the corrections to the mass and tension,
and this paper derives the eccentricity and ``Archimedes
effect''. Moreover, here we go beyond and demonstrate a method for an
arbitrary number of successive approximations.

\noindent {\bf Note added}: In the 3rd version we performed some
rewriting of subsection \ref{monopole-subsection} where we made
minor corrections and improved the clarity and we corrected a
trivial factor of 2 in (\ref{y1}).

\section{The asymptotic zone}
\label{newtoniansec}

In this section we write the static Einstein equations in a form that
will be convenient for iterative expansion in a small parameter around
the flat Minkowsky spacetime. This type of expansion is known as
``post-newtonian expansion" (see~\cite{Damour}) and we follow here the
usual conventions for this type of expansion. The leading order in the
expansion is the newtonian approximation. We repeat here the
calculation of the newtonian approximation for the caged black hole
which appeared in many places (see for example~\cite{HO1,H4}). The
following orders in the expansion are called ``post-newtonian" and
their calculation will appear in~\cite{GKin-progress}.

For the post-newtonian expansion it is convenient to write the Ricci
tensor in the following form~\cite{fock2}
\begin{equation}
R_{\mu\nu}=-\frac{1}{2}g_{\mu\sigma}g_{\nu\rho}g^{\alpha\beta}
\frac{\partial^{2}g^{\rho\sigma}}{\partial x^{\alpha}\partial
x^{\beta}}+\Gamma_{\mu}^{\alpha\beta}\Gamma_{\nu,\alpha\beta}-\Gamma_{\mu\nu},
\label{posteq}
\end{equation}
where $\Gamma_{\mu,\alpha\beta}$ and $\Gamma_{\mu}^{\alpha\beta}$ are
the Christoffel symbols of the first and the second kind,
respectively, and in addition one defines
\begin{eqnarray}
\Gamma_{\mu\nu}&\equiv&
\frac{1}{2}\bigg(g_{\mu\rho}\frac{\partial\Gamma^{\rho}}{\partial
x^{\nu}}+ g_{\nu\sigma}\frac{\partial\Gamma^{\sigma}}{\partial
x^{\mu}}- g_{\mu\rho}g_{\nu\sigma}\frac{\partial
g^{\rho\sigma}}{\partial x^{\alpha}}\Gamma^{\alpha} \bigg) \,,
\nonumber\\
\Gamma^{\nu} &\equiv& g^{\alpha\beta}\Gamma^{\nu}_{\alpha\beta}\,.
\nonumber
\end{eqnarray}
Next one chooses the harmonic (or de Donder~\footnote{The first
  introduction of this gauge appeared in~\cite{deDon}.}) gauge by the
requirement that
\begin{equation} \label{gaugeE}
\Gamma^{\nu}=\Box
x^{\nu}=\frac{1}{\sqrt{-g}}\frac{\partial}{\partial
x^{\beta}}(\sqrt{-g} g^{\beta\nu}) \equiv 0\,.
\end{equation}
where we denote by $g$ the determinant of the metric $g_{\mu\nu}$.  In
this gauge, the last term in the expression of the Ricci tensor above
vanishes. This choice of gauge is very convenient for expansion in the
asymptotic zone. Finally one attempts to solve Einstein's equations.

The first step in this iterative procedure is to look at the
linearized equations valid for weakly gravitating regions, namely
making the newtonian approximation. The metric is taken to be
\begin{equation}
g_{\mu\nu} = \eta_{\mu\nu}+h_{\mu\nu}\,.
\end{equation}
The harmonic gauge equation~(\ref{gaugeE}) takes the more famous form
(for example in the treatment of gravitational waves)
\[ \left(h_{\mu\nu}-\frac{1}{2}  h^{\alpha}_{\alpha}\,
\eta^{\mu\nu}\right)_{,\nu}=0\,.\]
One defines
\begin{equation}
\bar{h}_{\mu\nu}\equiv h_{\mu\nu}-\frac{1}{2}
h^{\alpha}_{\alpha}\,\eta_{\mu\nu}\,,
\label{hbar-h}
\end{equation}
in terms of which the linearized field equations become
\begin{equation}
\frac{1}{2}\, \Box \bar{h}_{\mu\nu} = G_{\mu\nu} = 8
 \pi\, G_d\, T_{\mu\nu} \,,
\end{equation}
where $\Box$ is the flat space D'alambertian.

\pagebreak[3]

In our case, working on the covering space implies an infinite array
of newtonian sources in the $z$ direction and the only non-zero
components of the energy momentum tensor is
\[
T_{00}=\sum_{n=-\infty}^{\infty} M\,\delta^{d-2}(x_1,\dots,x_{d-2})\,
\delta(z-nL) \,,
\]
where $(x_1,\dots,x_{d-2})$ denote the extended spatial
coordinates. The method of images can be used to solve the equation
for $\bar{h}_{00}$
\begin{eqnarray}
\bar{h}_{00}=-{\Phi_N \over 4 \pi} &\equiv& \frac{16\,\pi\,
G_{d}\,M}{(d-3)\, \Omega_{d-2}}\sum_{n=-\infty}^{\infty}
\frac{1}{\left(r^{2}+(z+nL)^{2}\right)^{\frac{d-3}{2}}}=
\nonumber\\
&=&{d-2 \over d-3}\, \rho_{0}^{d-3}\,
\sum_{n=-\infty}^{\infty}
\frac{1}{\left(r^{2}+(z+nL)^{2}\right)^{\frac{d-3}{2}}}\,,
\label{Newtonian-potential}
\end{eqnarray}
where $\Phi_N$ is the newtonian potential, conventionally normalized
such that its flux through a surface enclosing a mass $M$ is $4 \pi\,
G_d\, M$, $\rho_{0}$ is given by~\cite{myers}
\begin{equation}
\rho_{0}^{d-3} = \frac{16\,\pi\, G_{d}\,M}{(d-2)\,\Omega_{d-2}} \,,
\label{rho-m}
\end{equation}
and
\[\Omega_{d-2}=\frac{2\,\pi^{\frac{d-1}{2}}}{\Gamma(\frac{d-1}{2})}\,,\]
is the area of a unit $S^{d-2}$.  A more formal alternative to obtain
the prefactor of the newtonian potential in
equation~(\ref{Newtonian-potential}) is through matching with the
\Schw\ metric in the near zone.

We see that the first correction to the metric in the asymptotic
region is of order $\rho_0^{~d-3}$. We will see later that we must
choose the perturbation parameter in this region to be $\rho_0$ rather
than $\rho_0^{~d-3}$, namely\footnote{or more precisely $\rho_0$ to
  the power $gcd(2,d-3)$.}
\begin{equation}
g_{\mu\nu} = \sum_{n=0}^{\infty} \rho_0^{~n}\, g_{\mu\nu}^{(n)}\,,
\end{equation}
and hence we see that the leading correction comes at order $d-3$,
namely
\begin{eqnarray}
g^{(0)}_{\mu\nu} &=& \eta_{\mu\nu}\,,
\nonumber\\
g^{(d-3)}_{\mu\nu} &=& h_{\mu\nu} \,.
\end{eqnarray}

Transforming back to $h_{\mu\nu}$ using the inverse of~(\ref{hbar-h})
\begin{equation}
h_{\mu\nu} \equiv \bar{h}_{\mu\nu} -\frac{1}{d-2} \bar{h}^{\alpha}_{\alpha}\,
\eta_{\mu\nu}\,,
\label{h-hbar}
\end{equation}
yields the expression for the metric perturbation in terms of the
newtonian potential~(\ref{Newtonian-potential})
\begin{eqnarray}
h_{00} &=& -{d-3 \over d-2} {\Phi_N \over 4 \pi}
 = \rho_{0}^{d-3}\sum_{n=-\infty}^{\infty}\frac{1}{\left(r^{2}
+(z+nL)^{2}\right)^{\frac{d-3}{2}}},
\nonumber\\
h_{ij} &=& -{1 \over d-2} {\Phi_N \over 4 \pi}\, \delta_{ij}
= {1 \over d-3}\, h_{00}\,\delta_{ij}\,,
\label{Newtonian-metric}
\end{eqnarray}
where the Latin indices stand for the spatial components.

\pagebreak[3]

In the 5d case one can express the newtonian potential as
\begin{equation}
\Phi=-\frac{G_5\, M}{L\,r}\cdot\frac{\sinh\left(\frac{2\,\pi\,r}{L}\right)}
{\cosh\left(\frac{2\,\pi\,r}{L}\right)-\cos\left(\frac{2\,\pi\,z}{L}\right)}\,.
\end{equation}
In figure~\ref{black2} we give the equipotential surfaces of the
newtonian potential in 5d, and they look qualitatively the same in any
dimension $d>5$.

\FIGURE{\epsfig{file=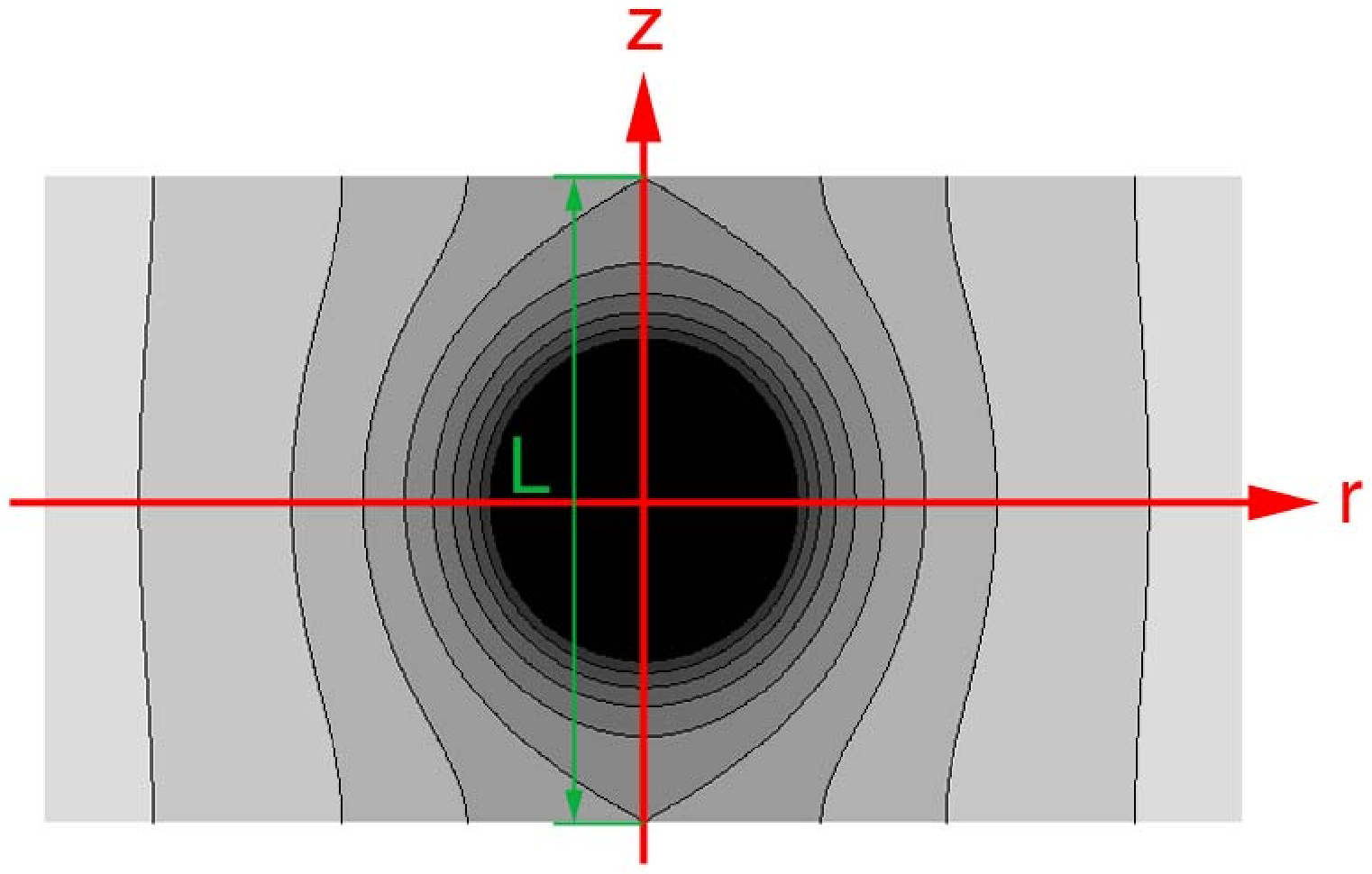, width=.5\textwidth, clip=}%
\caption{Equipotential surfaces of the newtonian potential in
  5d.\label{black2}}}

{\sloppy We close this section with two comments. First, at higher orders in
the perturbation procedure the form of the equations is dominated by
the linearized equations and is given by
\[\Box g^{(m),\mu\nu}=\it{F}(g,\partial g)\,,\]
where $(m)$ is the order under study and $\it{F}(g,\partial g)$ are
source terms which are quadratic, at least, in lower order metric
components and their derivatives. The second remark is that in this
section all the expressions for the metric components were in the
harmonic gauge. In the next sections we use the Schwarzschild gauge in
the near zone. To avoid cluttering the notation we will not introduce
always a different letter for every gauge --- in
subsection~\ref{trans} we give different notation for $\rho$ in the
two gauges ($\rho$ in the Schwarzschild coordinates and $\rho_{h}$ in
the harmonic gauge) but later we omit the difference in the
notation. However, it is important to remember the difference in the
gauge between the two zones.

}

\section{Black hole perturbations}
\label{BHperturb}

As explained in the introduction the zeroth order in the near
(horizon) zone is the $d$-dimensional Schwarzschild-Tangherlini
metric~\cite{tangher}
\[ ds^{2}=-f(\rho)\,dt^{2}+\frac{1}{f(\rho)}\,d\rho^{2}+\rho^{2}\,d
\Omega_{d-2}^{2}\,,\]
where $f(\rho)=1-\frac{\rho_{0}^{d-3}}{\rho^{d-3}}$, $\rho_0$ is
related to $M$ via~(\ref{rho-m}) and
\[d\Omega_{d-2}^{2}=d\chi^{2}+\sin^{2}\chi \,d \theta_{1}^{2}+\cdots +(\sin^{2}\chi\,\sin^{2}\theta_{1}\ldots \sin^{2}\theta_{d-4})\,d\theta_{d-3}^{2}\,,\]
is the metric on $S^{d-2}$.

\pagebreak[3]

{\sloppy In this section we find the linear static perturbations for the
$d$-dimensional Schwarzschild solution. Regge and Wheeler~\cite{Regge}
derived the linear equations that describe small perturbations to the
four dimensional Schwarzschild black hole, and here we generalize
their method for the static case.  These perturbations can be
interpreted as deviations of the black hole from spherical symmetry
due to remote masses. In the case of a compact dimension, we are
interested in the influence of the black hole (images) on
itself. Therefore we assume that the $\SO(d-2)$ symmetry in the
spherical coordinates $(\theta_{1},\ldots ,\theta_{d-3})$ is still
preserved and the deformation of the black hole takes place only in
the $(\rho,\chi)$ plane (which in ``cylindrical'' coordinates will be
part of the $(r,z)$ plane). We denote in this section the
$d$-dimensional Schwarzschild metric by $g_{\mu\nu}$ and the
perturbation metric by $h_{\mu\nu}$. Therefore, $h_{\mu\nu}$ is a
function only of $\rho$ and $\chi$.

}

The linearized vacuum Einstein equations can be brought to the
simplified form~\cite{wald}
\begin{equation}
\label{linear}
\delta R_{\mu\nu}=-\frac{1}{2}D_{\mu}D_{\nu}h-\frac{1}{2}
D^{\sigma}D_{\sigma}h_{\mu\nu}+D^{\sigma}D_{(\nu}h_{\mu)\sigma}=0\,,
\end{equation}
where $h=g^{\mu\nu}h_{\mu\nu}$, $D^{\mu}$ is the covariant derivative
with respect to the background metric $g_{\mu\nu}$ and
$A_{(\alpha\beta)} \equiv \( A_{\alpha\beta} + A_{\beta \alpha}\)/2$
stands for the symmetric part of a tensor $A_{\alpha\beta}$.

\subsection{Spherical harmonics on $\IS^{d-2}$}

Our goal is to simplify the equations by reducing them to a system of
ordinary differential equations. Following Regge and Wheeler we start
by expanding the solution, $h_{\mu\nu}$, into generalized spherical
harmonics on the sphere $S^{d-2}$. Each component of $h_{\mu\nu}$ is
transformed under local coordinate changes of $S^{d-2}$ like a scalar,
a vector or a tensor. Hence, we decompose $h_{\mu\nu}$ into 3 types of
spherical harmonics: scalar, vector and tensor harmonics. The scalar
components are: $h_{tt}$, $h_{\rho\rho}$ and $h_{\rho t}$. The vectors
are: $(h_{t\chi},h_{t\theta_{1}},\ldots ,h_{t\theta_{d-3}})$ and
$(h_{\rho\chi},h_{\rho\theta_{1}},\ldots ,h_{\rho\theta_{d-3}})$. The
tensor is formed of the block
\[\left( \begin{array}{ccccc}
 h_{\chi\chi} & h_{\chi\theta_{1}} & h_{\chi\theta_{2}} & \cdots  & h_{\chi\theta_{d-3}} \\
  ** & h_{\theta_{1}\theta_{1}} & h_{\theta_{1}\theta_{2}} &\cdots  & h_{\theta_{1}\theta_{d-3}} \\
  ** & * & h_{\theta_{2}\theta_{2}} & \cdots & h_{\theta_{2}\theta_{d-3}} \\
  \vdots & \vdots & \vdots & \ddots & \vdots \\
  ** & * & * & \cdots & h_{\theta_{d-3}\theta_{d-3}}
\end{array}\right),
\]
where we denote by $*$ symmetric components. Counting degrees of
freedom we find that we should have 3 elements in the basis of the
scalar harmonics, $2\,(d-2)$ in the basis of the vector harmonics and
$\frac{(d-2)\,(d-1)}{2}$ in the basis of the tensor
harmonics. Together we obtain all the $\frac{d\,(d+1)}{2}$ components
of $h_{\mu\nu}$.

Since we are interested in expansion to harmonics which are static and
have $\SO(d-2)$ symmetry the non-vanishing components of the tensor
$h_{\mu\nu}$ are:
\begin{equation} \left( \begin{array}{cccccc}
  h_{tt} & & & &  &\\
  & h_{\rho\rho} & h_{\rho\chi} & & & \\
  & * & h_{\chi\chi} &  & & \\
  &  &  & h_{\theta_{1}\theta_{1}} & & \\
  & & & & \ddots & \\
  & & & & & h_{\theta_{d-3}\theta_{d-3}}
\end{array}
\right), \label{5ansatz}
\end{equation}
where in addition, the $\SO(d-2)$ symmetry implies that the only
independent angular components on the diagonal are $h_{\chi \chi}$ and
$h_{\theta_{1}\theta_{1}}$.  The rest of the angular components are
obtained through the relations
\[
h_{\theta_{i}\theta_{i}}=\sin^{2}{\theta_{i-1}}\,
h_{\theta_{i-1}\theta_{i-1}}\,,
\qquad i=2,\ldots ,d-3\,.
\]

Note that there are only 5 independent components of $h_{\mu\nu}$,
namely $h_{tt}$, $h_{\rho \rho}$, $h_{\rho \chi}$, $h_{\chi \chi}$,
$h_{\theta_{1}\theta_{1}}$, comprising of 2 scalars, 1 vector and 2
tensors, and these numbers are independent of the dimension.
Accordingly, we will find 5 linear ordinary differential equations
(the equations of motion).

\subsubsection{Scalar harmonics}

By considering the flat $(d-1)$-dimensional Laplace equation we can
get both the scalar spherical harmonics on $\IS^{d-2}$, and the
leading radial profile of the multipoles in the nearly flat region
$\rho \gg \rho_0$. Since we assume $\SO(d-2)$ symmetry we consider a
Laplace equation for a function which depends only on one angular
variable $\chi$. Thus, the Laplace equation for a function
$\Psi(\rho,\chi)$ on $\IR^{d-1}$ would be
\begin{equation}
\partial_{\rho} \( \rho^{d-2}\,\sin^{(d-3)}(\chi)\,\partial_{\rho}
\Psi(\rho,\chi) \) + \rho^{d-4}\, \partial_{\chi} \(
\sin^{(d-3)}(\chi)\,\partial_{\chi}\Psi(\rho,\chi) \) =0\,.
\end{equation}
Separation of variables $\Psi(\rho,\chi)=R(\rho)\,\Pi(\chi)$ gives us
two separated equations for each eigenvalue $l$ which denotes also the
number of the multipole in the expansion
\[
\Psi(\rho,\chi)=\sum_{l=0}^{\infty}R_{l}(\rho)\,\Pi^{l,d}_{0}(\chi)\,.
\]

$\Pi^{l,d}_{0}(\chi)$ is the angular function which is associated with
each one of the multipoles. The extra zero index in the angular
function stands to remind us that had we not had the $\SO(d-2)$
symmetry, the angular functions, which depends on $\chi$, would have
additional indices (additional ``quantum numbers"). (For an example
see the 5d case later in this subsection). The equation for
$\Pi^{l,d}_{0}(\chi)$ is
\begin{equation}  \label{pieq}
\frac{d^{2}\,\Pi^{l,d}_{0}}{d\chi^{2}}+(d-3)\,\cot(\chi)
\frac{d\,\Pi^{l,d}_{0}}{d\chi}+l\,(l+d-3)\,\Pi^{l,d}_{0}=0\,,
\end{equation}
where $l\,(l+d-3)$ are the eigenvalues. This equation can be brought
to the form of a Legendre equation~\cite{Muller} in $d-1$ dimensions
using the substitution $t=\cos(\chi)$ (then the solutions of the
equation are called Legendre polynomials of the variable $t$ in $d-1$
dimensions). The solutions can be expressed by a Rodriguez formula
(see~\cite{Muller})
\begin{equation} \label{rodrig}
\Pi^{l,d}_{0}(\chi)=\frac{\Gamma(\frac{d}{2}-1)}{2^{l}\,\Gamma(l+\frac{d}{2}-1)}\,
\sin^{4-d}(\chi)\,\left(\frac{1}{\sin(\chi)}\,\frac{d}{d\chi}\right)^{l}\,
\sin^{2l+d-4}(\chi)\,,
\end{equation}
where the prefactor with the Gamma functions fixes the usual
normalization of the Legendre polynomials.

The functions $R_{l}(\rho)$ give us the radial part of the
expansion. $R_{l}(\rho)$ is obtained as the solution of the eigenvalue
equation
\begin{equation}
\frac{d}{d\rho} \( \rho^{d-2}\,R_{l}\,'(\rho) \)
-l\,(l+d-3)\,\rho^{d-4}\,R_{l}(\rho)=0\,.
\end{equation}
Therefore
\begin{equation} \label{multilap}
R_{l}(\rho)=J_{l}\,\rho^{-(l+d-3)}+K_{l}\,\rho^{l}\,,
\end{equation}
where $J_{l}$ and $K_{l}$ are constants. For fixed $l$ and $d$ the
first term is the multipole of a mass distribution at $\rho=0$ and the
second term is the multipole of a mass distribution at infinity. Note
that $l=0$ is the first multipole --- the monopole, $l=1$ is the
dipole and so on.

\pagebreak[3]

Since the linear equations for $h_{\mu \nu}$~(\ref{linear}) are
$\SO(d-1)$ invariant we may separate the angular variables, and since
the perturbed metric depends only on $(\rho,\chi)$, we may expand the
scalar components $h_{tt}$ and $h_{\rho\rho}$ in spherical harmonics
as follows
\begin{eqnarray} h_{tt} &=&
  \sum_{l=0}^{\infty}\tilde{A_{l}}(\rho)\,\Pi^{l,d}_{0}(\chi)\,,
\nonumber\\
h_{\rho\rho} &=&
  \sum_{l=0}^{\infty}\tilde{B_{l}}(\rho)\,\Pi^{l,d}_{0}(\chi) \,,
  \label{scalar}
\end{eqnarray}
where the radial functions $\tilde{A_{l}}(\rho),\,
\tilde{B_{l}}(\rho)$ satisfy some differential equations to be
discussed later.

\paragraph{5d scalar spherical harmonics.}

As a concrete simple example, where we can give explicit formulae for
the scalar harmonics, let us consider the 5d case. Gerlach and
Sengupta used this type of decomposition in 5d for the
Robertson-Walker spacetime~\cite{gerlach}. Let us denote the scalar
spherical harmonics in 5 dimensions\footnote{Note that in 5 dimensions
  we use different notation for the indices; we keep the index $l$ for
  the usual spherical harmonics $Y_{lm}(\theta,\varphi)$. Thus, we
  will use $\Pi^{n,5}_{0}$ instead of $\Pi^{l,d}_{0}$ in the general
  case.}  by $Q^{n}_{lm}(\chi,\theta,\varphi)$. They can be separated
into a product of two types of functions
\[Q^{n}_{lm}(\chi,\theta,\varphi)=\Pi^{n,5}_{l}(\chi)\,Y_{lm}
(\theta,\varphi)\,,\]
where $Y_{lm}(\theta,\varphi)$ are the usual spherical harmonics on
$S^{2}$ and $\Pi^{n,5}_{l}(\chi)$ are the ``Fock" harmonics
(see~\cite{gerlach,fock}), which are given by
\[\Pi^{n,5}_{l}(\chi)=\sin^{l}(\chi)\,\frac{d^{l+1}\,(\cos((n+1)\,\chi))}{d(\cos(\chi))^{l+1}}\,.\]

Since we require $\SO(3)$ symmetry for the $(\theta,\varphi)$
two-sphere, we take $l=m=0$ in the spherical harmonics. Thus we arrive
to~(\ref{pieq}) in the 5d case (with the index $n$ instead of $l$)
\begin{equation}
\frac{d^{2}\,\Pi^{n,5}_{0}}{d\chi^{2}}+2\,\cot(\chi)\,\frac{d\,\Pi^{n,5}_{0}}{d\chi}+n\,(n+2)\,\Pi^{n,5}_{0}=0\,,
\end{equation}
whose solutions are Chebyshev polynomials of the second kind
\[\Pi^{n,5}_{0}(\chi)=\frac{\sin[(n+1)\,\chi]}{\sin(\chi)} \equiv U_{n}(\cos(\chi))\,.\]

\subsubsection{Vector and tensor harmonics}

Given a family of scalar harmonics one can form a family of ``scalar
derived'' vector harmonics simply by taking its gradient.  There are
other, more involved, vector harmonics as well, but we shall see now
that the ``scalar derived'' family suffices for our purposes. For a
concrete example of a basis of vector harmonics in 5d (on $\IS^{3}$)
see appendix~\ref{ap C}.

Due to the symmetries the vector has a single
component~(\ref{5ansatz})
\begin{equation}
 h_{\rho \mu} = h_{\rho \chi}(\rho,\chi)\, \delta_{\mu \chi}\,,
\end{equation}
where $\mu$ runs over the angular coordinates only. Denoting $\hat{h}
(\rho,\chi)\equiv \int^\chi d\chi'\, h_{\rho \chi}(\rho,\chi')$ we
have
\begin{equation}
 h_{\rho \mu} = \del_\mu \hat{h} (\rho,\chi) \,.
\label{h-from-hat-h}
\end{equation}

Therefore, after expanding $\hat{h}$ into spherical harmonics and
substituting in~(\ref{h-from-hat-h}) we can write the expansion of
$h_{\rho\chi}$ into radial functions $\tilde{C_{l}}(\rho)$ as
\begin{equation}
h_{\rho\chi}=\sum_{l=0}^{\infty} \tilde{C_{l}}(\rho)\, \del_\chi
\Pi^{l,d}_{0}(\chi)\,.
\label{vector}
\end{equation}

A similar argument holds for the tensor components which are
essentially $h_{\chi \chi}$, \linebreak
$h_{\theta_1,\theta_1}$~(\ref{5ansatz}). Again there are exactly two
``scalar derived'' tensor harmonics
 \begin{equation}
  \hat{D_{i}} \hat{D_{j}} \Pi^{l,d}_{0}\,,
\end{equation}
and
\begin{equation}
\gamma_{ij}\,\Pi^{l,d}_{0}\,,
\end{equation}
where $\gamma_{ij}$ is the metric on the sphere $S^{d-2}$,
$\hat{D_{i}}$ is the covariant derivative on $S^{d-2}$, and the second
term is proportional to the trace of the first. So we have two tensor
harmonics to decompose into, which is exactly the number we need, and
indeed one can verify that these two families always suffice.

Hence we can decompose the tensor part into radial functions
$\tilde{D_{l}}(\rho)$, $\tilde{E_{l}}(\rho)$
\begin{eqnarray}
&& \sum_{l=0}^{\infty} \tilde{D_{l}}(\rho) \left(
\begin{array}{cccc}
 \frac{ \partial^{2}}{ \partial
\chi^{2}} & &  &\\
  & \sin(\chi)\cos(\chi)  \frac{\partial}{ \partial
\chi}& 0 &\\
  &  &  \sin(\chi)\cos(\chi)\sin^{2}(\theta_{1})  \frac{\partial}{ \partial
\chi}&\\
 & & & \ddots
\end{array}
\right)\Pi^{l,d}_{0}(\chi) +
\nonumber\\
&& + \sum_{l=0}^{\infty}
\tilde{E_{l}}(\rho) \left( \begin{array}{cccc}
 1 &  &  &\\
   & \sin^{2}(\chi) &  &  \\
   &  & \sin^{2}(\chi)\sin^{2}(\theta_{1}) & \\
   & & & \ddots
\end{array}
\right)\Pi^{l,d}_{0}(\chi)\,.
\label{tensor}
\end{eqnarray}

The discussion above led us to the conclusion that the only elements
of both the vector and tensor basis which survived under our symmetry
requirements are the ``scalar derived'' ones.  These are in the
``scalar type'' representation under the $\SO(d-1)$ rotation group,
where by ``scalar type'' we mean representations whose Dynkin indices
are $[l,0,\dots,0]$ for some $l \ge 0$, namely those in the $l$-times
traceless symmetric product of the vector representation. More
generally, Kodama and Sasaki~\cite{kodama} pointed out that one can
classify the different basis elements into three groups with respect
to their different representations under the rotation isometry group:
scalar, vector and tensor ``type''.\footnote{The vector and tensor
  types are defined in analogy with the scalar type: a vector type
  representation has Dynkin indices $[l,1,0,\dots,0]$, namely the
  traceless product of the 2nd rank antisymmetric representation with
  the $l$-times traceless symmetric product of the vector
  representation, and a tensor type is $[l,2,0,\dots,0]$.}  This
classification into representations of the isometry group should not
be confused with the classification that we used above with respect to
local coordinate transformations of $S^{d-2}$. So, according to this
classification,~(\ref{vector}) is a vector of ``scalar type'' under
the rotation group and~(\ref{tensor}) is a tensor of ``scalar type''
as well. Since we deal with linear equations two different
representations cannot mix. Thus even if there were any
representations of non-scalar ``type'' in the decomposition of the
perturbed metric, they would not appear in the equations for the
scalar type radial functions $\tilde{A_{l}},\, \tilde{B_{l}},\,
\tilde{C_{l}},\, \tilde{D_{l}},\, \tilde{E_{l}}$.

\subsection{The choice of gauge}

We have now 5 radial fields
($\tilde{A_{l}}(\rho)$,..,$\tilde{E_{l}}(\rho)$) defined
in~(\ref{scalar}),~(\ref{vector}),~(\ref{tensor}) for each mode of the
expansion. We can reduce further the number of fields to 3 using the
gauge freedom. The gauge transformations of linearized general
relativity about a solution $g_{\mu\nu}$ are of the form
\[h_{\mu\nu}\longrightarrow h_{\mu\nu}+D_{\nu}\xi_{\mu}+D_{\mu}\xi_{\nu}\,,\]
where $\xi^{\mu}$ is an arbitrary vector field --- the generator of
an infinitesimal transformation. The most general generator
consistent with the symmetries is
\begin{eqnarray}
\xi_{t}&=& 0\,,
\nonumber\\
\xi_{\rho}&=&\sum_{l=0}^{\infty}L_{l}(\rho)\,\Pi^{l,d}_{0}(\chi)\,,
\nonumber\\
\xi_{\chi}&=&\sum_{l=0}^{\infty}M_{l}(\rho)\,\Pi^{l,d \prime}_{0}(\chi)\,,
\nonumber\\
 \xi_{\theta_{1}} &=& 0, \ldots ,  \xi_{\theta_{d-3}}=0\,.
\end{eqnarray}
It is natural to eliminate the ``scalar derived'' functions
$\tilde{C_{l}}(\rho)$ and $\tilde{D_{l}}(\rho)$ in each mode, putting
$h_{\mu\nu}$ in a diagonal form. We term this the {\it ``no derivative
  gauge''}. Thus,
\begin{eqnarray*}
h_{\rho\chi} &\longrightarrow&
h_{\rho\chi}+D_{\chi}\xi_{\rho}+D_{\rho}\xi_{\chi}
\\
&=&h_{\rho\chi}+\xi_{\rho,\chi}+\xi_{\chi,\rho}-2\,\Gamma_{\rho\chi}^{\chi}\,\xi_{\chi}
\\
&=&\sum_{l=0}^{\infty}\left(\tilde{C_{l}}+L_{l}+M_{l}'-\frac{2}{\rho}\,M_{l}\right)\,\Pi^{l,d \prime}_{0}(\chi)\,,
\\
h_{\chi\chi} &\longrightarrow& h_{\chi\chi}+2\,D_{\chi}\xi_{\chi}
\\
&=&h_{\chi\chi}+2\,\xi_{\chi,\chi}-2\,\Gamma_{\chi\chi}^{\rho}\,\xi_{\rho}
\\
&=&\sum_{l=0}^{\infty}\left(\tilde{E_{l}}+2\,\rho \,f \,L_{l}\right)\,\Pi^{l,d}_{0}(\chi)+\left(\tilde{D_{l}}+2\,M_{l}\right)\Pi^{l,d \prime\prime}_{0}(\chi)\,.
\end{eqnarray*}
The required gauge implies that the functions $L_{l}(\rho)$ and
$M_{l}(\rho)$ should satisfy the following conditions
\begin{eqnarray}
\tilde{C_{l}}+L_{l}+M_{l}'-\frac{2}{\rho}\,M_{l}&=&0\,,
\nonumber \\
\tilde{D_{l}}+2\,M_{l}&=&0\,.
\end{eqnarray}
Moreover, since $\tilde{E_{l}}$ is arbitrary function of $\rho$ we can
redefine it using the transformation $\tilde{E_{l}}\rightarrow
\tilde{E_{l}}+2\,\rho\, f\, L_{l}$. These equations for $L_{l}(\rho)$
and $M_{l}(\rho)$, although including differential operators, are
actually algebraic and have a single solution (for each mode) without
any additional gauge freedom, i.e., \emph{the gauge is completely
  fixed}. Applying the gauge transformation to any component of
($h_{\theta_{1}\theta_{1}}$,\ldots ,$h_{\theta_{d-3}\theta_{d-3}}$)
yields the same equations as in the case of $h_{\chi\chi}$
using~(\ref{pieq}).

\subsection{The field equations}
\subsubsection{The master equation}

After gauge fixing the perturbed metric diagonalizes and we are
left with 3 metric functions $h_{tt},\, h_{\rho \rho},\, h_{\chi
\chi}$, which are functions of $(\rho,\chi)$
\begin{eqnarray}
 h_{\mu \nu}\, dx^\mu\, dx^\nu &=&
 \tilde{A}(\rho,\chi)\, dt^2 + \tilde{B}(\rho,\chi)\, d\rho^2 +
 \tilde{E}(\rho,\chi)\, d\Omega_{d-2}^{~2}
\nonumber\\
 &=& \sum_{l=0}^{\infty} \( \tilde{A_{l}}(\rho)\, dt^2
+ \tilde{B_{l}}(\rho)\, d\rho^2 + \tilde{E_{l}}(\rho)
 d\Omega_{d-2}^{~2} \) \Pi^{l,d}_{0}(\chi)\,.
\label{3ansatz}
\end{eqnarray}

It turns out that Einstein's equations simplify if we use
$h^\mu_{~\nu}$ instead of $h_{\mu \nu}$. Thus we define $A \equiv
-\frac{\tilde{A}}{f}$, $ B \equiv f\, \tilde{B}$, $ E \equiv
\frac{\tilde{E}}{\rho^{2}}$ where as usual
$f(\rho)=1-\frac{\rho_{0}^{d-3}}{\rho^{d-3}}$. The ansatz in the new
variables reads
\begin{eqnarray}
h^\mu_{~\nu}&=& \left(
\begin{array}{ccccc}
 A(\rho,\chi) &  &  &  &  \\
 & B(\rho,\chi) &  &  &  \\
 &  &   E(\rho,\chi) &  &\\
 &     &  & \ddots & \\
 & &  & & E(\rho,\chi)\ldots
\end{array}
\right)
\nonumber\\
&=& \sum_{l=0}^{\infty} \left(
\begin{array}{ccccc}
 A_{l}(\rho) &  &  &  &  \\
 & B_{l}(\rho) &  &  &  \\
 &  &   E_{l}(\rho) &  &\\
 &     &  & \ddots & \\
 & &  & & E_{l}(\rho)\ldots
\end{array}
\right) \Pi^{l,d}_{0}(\chi)\,.
\label{3ansatzb}
\end{eqnarray}

We substitute this ansatz into the linearized Einstein
equations,~(\ref{linear}) and using~(\ref{pieq}) we obtain, as
expected, five equations
\paragraph{$\delta R_{tt}=0;$}
\begin{equation}
2\,l\,(l+d-3)\,A_{l}-(3\,\rho^{2}\,f'+2\,(d-2)\rho
\,f)\,A_{l}'+\rho^{2}\,f'\,B_{l}'
-(d-2)\,\rho^{2}\,f'\,E_{l}'-2\,\rho^{2}\,f\,A_{l}''=0\,,
\label{eq1}
\end{equation}
\paragraph{$\delta R_{\rho \rho}=0;$}
\begin{eqnarray}
2\,l\,(l+d-3)\,B_{l}
-3\,\rho^{2}\,f'\,A_{l}'+2\,(d-2)\,\rho
f\,{B}'_{l}+\rho^{2}\,f'\,B_{l}'-&&
\nonumber\\
-4\,(d-2)\,\rho\, f\,E_{l}'
-(d-2)\,\rho^{2}\,f'\,E_{l}'+2\,\rho^{2}\,f\,A_{l}''
-2\,(d-2)\,\rho^{2}\,f\,E_{l}''&=&0\,,
\label{eq2}
\end{eqnarray}
\paragraph{$\delta R_{\rho \chi}=0;$}
\begin{equation} \label{eq3}
 -(\rho\, f'-2\,f)\,A_{l}+2\,(d-3)\,f\,B_{l}+\rho
\,f'\,B_{l} - 2\,\rho\, f \,A_{l}'-2\,(d-3)\,\rho \,f\,E_{l}'=0\,,
\end{equation}
\paragraph{$\delta R_{\chi \chi}=0;$}
\begin{equation}
\label{eq4}
-b_{l}(\rho)\,\Pi^{l,d}_{0}(\chi)-\frac{1}{2}\,a_{l}(\rho)\,
\Pi^{l,d\prime\prime}_{0}(\chi)=0\,,
\end{equation}
\paragraph{$\frac{\delta R_{\theta \theta}}{\sin^{2}(\chi)}-\delta R_{\chi \chi}=0;$}
\begin{equation} \label{eq5}
-\frac{1}{2}\,
a_{l}(\rho)\,\left(\cot(\chi)\,\Pi^{l,d\prime}_{0}(\chi)-\Pi^{l,d\prime\prime}_{0}(\chi)\right)=0\,,
\end{equation}
where we define
\begin{eqnarray}
a(\rho) &\equiv& A +B +(d-4)\,E = \mbox{tr}(h)-2 E\,,
\nonumber\\
a(\rho) &=& \sum_{l=0}^{\infty}\, a_{l}(\rho)\, \Pi^{l,d}_{0}(\chi)\,,
\\
b_{l}(\rho) &\equiv& (d-3)\,(E_{l}-B_{l})+\frac{1}{2}\,\rho \,f\,(
A_{l}^{\prime}-B_{l}^{\prime})+\frac{1}{2}\,\rho^{2}\,
f^{\prime}\,(E_{l}^{\prime}+E_{l}^{\prime\prime})+
\nonumber\\
&& +(d-2)\,\rho \,f \,E_{l}^{\prime}-\frac{1}{2}\,l\,(l+d-3)\,E_{l}\,.
\end{eqnarray}

Three of the equations are second order in the derivatives. These are
the evolution equations. One of the equations is first order ($\delta
R_{\rho \chi}=0$) --- the constraint equation. For $l>0$ the
expression in the brackets in~(\ref{eq5}) is non-zero and we get the
following algebraic relation
\begin{equation}
 \fbox{$\room \displaystyle  a_{l}(\rho)=A_{l}+B_{l}+(d-4)\,E_{l}=0\,.$}
\label{algeb}
\end{equation}
The case $l=0$ is degenerate and we will discuss it separately.

Now the variables can be separated in~(\ref{eq4}) and it becomes
a second order ordinary differential equation
\begin{eqnarray}
b_{l}(\rho) &=& (d-3)\,(E_{l}-B_{l})
+\frac{1}{2}\,\rho \,f\,(A_{l}^{\prime}-B_{l}^{\prime})
+\frac{1}{2}\,\rho^{2}\,f^{\prime}\,(E_{l}^{\prime}+E_{l}^{\prime\prime})+
\nonumber\\
&& +(d-2)\,\rho \, f\,E_{l}^{\prime}-\frac{1}{2}\,l\,(l+d-3)\,E_{l}=0\,.
 \label{eq6}
\end{eqnarray}

Using the algebraic relation to eliminate $B_{l}$ and its derivative
from the other equations, equations~(\ref{eq1}) and~(\ref{eq2}) become
\begin{eqnarray}
-c_{l}(\rho)+l\,(l+d-3)\,A_{l}-\rho^{2}\,f'\,(d-3)\,E_{l}'&=&0\,,
\nonumber\\
-c_{l}(\rho)-l\,(l+d-3)\,A_{l}-l\,(l+d-3)\,(d-4)\,E_{l}-&&
\nonumber\\
 -E_{l}'\,\left[(d-2)^{2}\,\rho
\,f+\rho^{2}\,f'\,(d-3)\right]-(d-2)\,\rho^{2}\,f\,E_{l}''&=&0\,,
\end{eqnarray}
where we use the abbreviation
\[c_{l}(\rho)\equiv (2\,\rho^{2}\,f'+(d-2)\,\rho\,f)\,A_{l}'+\rho^{2}\,f\,A_{l}''\,.\]
>From the last two equations we can express $A_{l}(\rho)$ in terms of
$E_{l}(\rho)$ and its derivatives
\begin{equation} \label{Aneq}
 -A_{l}=\frac{(d-2)\,\rho
\,f}{2\,l\,(l+d-3)}\,\left[(d-2)\,E_{l}'+\rho\,E_{l}''\right]+\frac{(d-4)
}{2}\,E_{l} \,.
\end{equation}
Now using the remaining equations ((\ref{eq3})--(\ref{eq4})), we arrive
to a second order linear differential equation for $E_{l}$ where we
introduce $x \equiv \frac{\rho}{\rho_{0}}$, a dimensionless variable
\begin{equation} \label{geneq}
\alpha_{l}(x)\,E_{l}(x)+\beta_{l}(x)\,E_{l}^{\prime}(x)
+\gamma_{l}(x)\,E_{l}^{\prime\prime}(x)=0 \,,
\end{equation}
where
\begin{eqnarray*}
\alpha_{l}(x)&\equiv& -l\,(l+d-3)\,x\,[(l+d-2)\,(l-1)\,x^{d-3}-(d-4)\,(d-2)]\,,
\\
\beta_{l}(x)&\equiv& (d-2)\,(l+d-2)\,(l-1)\,x^{d-1}-[l^{2}+(d-3)\,l-2\,(d-2)^{2}]\,x^{2}-(d-2)^{2}\,x^{5-d}\,,
\\
\gamma_{l}(x)&\equiv& (l-1)\,(l+d-2)\,x^{d}+[(d-2)-(l-1)\,(l+d-2)]\,x^{3} -(d-2)\,x^{6-d}\,.
\end{eqnarray*}

Summarizing, we reduced the equations of static ($l\geq1$, scalar
type) perturbations into a single second order ordinary equation for
the function $E_{l}$~(\ref{geneq})). This is our ``master equation''
in the variable $x$. The full solution for the perturbation metric
$h_{\mu\nu}$ can be constructed from the solution for $E_{l}$
using~(\ref{Aneq}) and~(\ref{algeb}).

\subsubsection{The monopole perturbations}
\label{monopole-subsection}

The solution for $l=0$ is equivalent according to Birkhoff's
theorem (to leading order) with a \Schw ~black hole  possibly with
different parameters (mass and asymptotic potential) but without a
change of its shape (higher multipoles).

In this case we have $\Pi^{0,d}_{0}(\chi)\equiv1$. Hence, the
algebraic relation is absent. In this case we have only three
equations
\paragraph{$\delta R_{tt}=0;$}
\begin{equation} \label{mono1}
-(3\,\rho^{2}\,f'+2\,(d-2)\,\rho\, f
)\,A_{0}^{\prime}+\rho^{2}\,f'\,B_{0}^{\prime}-(d-2)\,\rho^{2}\,
f'\,E_{0}^{\prime} -2\,\rho^{2}\,f\,A_{0}^{\prime\prime}=0\,,
\end{equation}
\paragraph{$ \delta R_{\rho\rho}=0;$}
\begin{eqnarray}
-3\,\rho^{2}\,f'\,A_{0}^{\prime}+2\,(d-2)\,\rho\, f\,B_{0}^{\prime}
 +\rho^{2}\,f'\,B_{0}^{\prime}-4\,(d-2)\,\rho\,
f\,E_{0}^{\prime}-&&
\nonumber\\
-(d-2)\,\rho^{2}\,f'\,E_{0}^{\prime}
-2\,\rho^{2}\,f\,A_{0}^{\prime\prime}-2\,(d-2)\,\rho^{2}\, f\,
E_{0}^{\prime\prime} &=&0\,,
\label{mono3}
\end{eqnarray}
\paragraph{$ \delta R_{\chi\chi}=0;$} (or $\delta R_{\theta\theta}=0$)
\begin{equation} \label{mono2eq}
 (d-3)\,(E_{0}-B_{0})+\frac{1}{2}\,\rho\, f\,(
A_{0}^{\prime}-B_{0}^{\prime})
 +\frac{1}{2}\,\rho^{2}\,f^{\prime}\,E_{0}^{\prime}+\frac{1}{2}\,\rho^{2}\,f\,E_{0}^{\prime\prime}
+(d-2)\,\rho\, f\, E_{0}^{\prime}=0\,.
\end{equation}

Subtracting~(\ref{mono1}) from~(\ref{mono3}) we obtain a simple
equation, where we use again the dimensionless variable
$x={\rho}/{\rho_{0}}$, for later convenience
\[A_{0}^{\prime}+B_{0}^{\prime}-2\,E_{0}^{\prime}-x\,E_{0}^{\prime\prime}=0\,.\]
The solution of this equation is the following relation between the
perturbation metric components
\[A_{0}+B_{0}-E_{0}-x\,E_{0}^{\prime}=-2\,C_{1}\,,\]
where $C_{1}$ is a constant. One can verify by substitution that
this is the solution of~(\ref{mono1}) as well, i.e.,~(\ref{mono1})
does not give us any additional restriction on the solution.
 Altogether we have 2 independent equations for the three fields
$A_0,\, B_0,\, E_0$. Namely, the solution is determined up to an
arbitrary function. This freedom is exactly a residual gauge
freedom expected of a \Schw ~solution $\rho \rightarrow
\overline{\rho}(\rho)$: it is a result of the unbroken $\SO(d-1)$
spherical symmetry.  Thus, the general solution should contain, in
addition to the free function, two constants corresponding to a
change in the mass, and an asymptotic potential (or equivalently,
a rescaling of $t$). One of the constants will be set by matching,
and the other one corresponds to a change of parametrization of
the branch of solutions, namely, a change of ``scheme''.


In analogy with the case $l\geq1$, we
 choose to
fix the gauge by the requirement that the algebraic condition will
be satisfied for the monopole as well, i.e., \begin{equation}
a_{0}(\rho)= A_{0}+B_{0}+(d-4)\,E_{0}=0\,. \label{monogauge}
\end{equation}
 This eliminates all the gauge freedom apart from a constant shift
of the radial coordinate, which will be a third constant in the
general solution.
Combining the previous two equations together, we obtain a simple
differential equation for $E_{0}$
\[(d-3)\,E_{0}+x\,E_{0}^{\prime}-2\,C_{1}=0\,.\]
The solution contains two constants which we denote by $C_{1}$ and
$C_{2}$
\[E_{0}(x)=\frac{2\,C_{1}}{d-3}+\frac{C_{2}}{x^{d-3}}\,.\]
Since we use the same algebraic condition as in the $l\geq1$ case, we
can obtain the same result using~(\ref{geneq}) for $l=0$.
  Substituting this result into
 (\ref{mono2eq},\ref{monogauge})
(and solving it), we obtain all the metric components for the
monopole perturbation
\begin{eqnarray}
h_{\chi\chi}^{(0),d}&=&\frac{2\,C_{1}}{d-3}\,\rho_{0}^{2}\,x^{2}+\frac{C_{2}\,\rho_{0}^{2}}{x^{d-5}}\,,
\label{monochi}\\
h_{tt}^{(0),d}&=&2\,C_{1}-\frac{C_{2}\,(d-3)}{2\,x^{2d-6}}+\frac{C_{3}}{x^{d-3}}\,,
\\
h_{\rho\rho}^{(0),d}&=&\frac{1}{f}\,\left[2\,C_{1}\,\left(\frac{1}{x^{d-3}\,f}+\frac{1}{d-3}\right)-\frac{C_{2}}{x^{d-3}}\left(\frac{d-5}{2}+\frac{d-3}{2\,f}\right)+\frac{C_{3}}{f\,x^{d-3}}\right],
\end{eqnarray}
where $C_{3}$ is an additional constant. We fix the gauge constant
by requiring that the location of the horizon remains $\rho_0$,
namely that at the horizon the coefficient of the singular second
order pole ($\propto \frac{1}{f^{2}}$) in $h_{\rho\rho}$ vanishes.
Therefore, we set 
\[C_{2}=\frac{2}{d-3}\,\left(2\,C_{1}+C_{3}\right).\]
Now, in addition, we choose $C_{3}$ 
 to be 
\[C_{3}=-2\,C_{1}\,.\]

These choices are
advantageous because now one can check that the solution to the
linearized equations (for the monopole) becomes an exact solution
of the Einstein equations
\begin{eqnarray}
h_{\chi\chi}^{(0),d}&=&\frac{2\,C_{1}\,\rho^{2}}{d-3}\,,
\label{allo1} \\
h_{tt}^{(0),d}&=&2\,C_{1}\,f\,,
\label{allo2}\\
h_{\rho\rho}^{(0),d}&=&\frac{2\,C_{1}}{(d-3)\,f}\,.
\label{allo3}
\end{eqnarray}
That is, the same form of solution is valid to any order of expansion
of Einstein equations in the near zone. In each order of the
perturbation only one constant, $C_{1}$, is needed to be determined
for the monopole correction and it will be done through the matching
process to be described in section~\ref{desc}.

\subsubsection{The solution of the master equation} \label{solution}

For $l\geq1$ let us make the following change of variable
in~(\ref{geneq})
\begin{equation} \label{theX}
X\equiv x^{d-3}\,,
\end{equation}
to obtain a Fuchsian equation with four regular singular
points\footnote{In the case $l=1$~(\ref{geneq}) becomes an equation
  with 3 regular singular points ($0$,$1$ and $\infty$), namely a
  Hypergeometric equation. The solution of this equation describes a
  translation of the black hole along the $z$ axis.}
\begin{equation}
\label{keq} \fbox{$ \room \displaystyle   \frac{d^{2}\,E_{l}}{dX^{2}}
+\left(\frac{2}{X}+\frac{1}{X-1}-\frac{1}{X-w_{l,d}}\right)\frac{d\,E_{l}}{dX}
-p_{l,d}\,(1+p_{l,d})\frac{X+(d-4)\,w_{l,d}}{X\,(X-1)\,(X-w_{l,d})}\,E_{l}=0\,,$} \label{master-eq}
\end{equation}
where
\[w_{l,d}\equiv-\frac{d-2}{(l-1)\,(l+d-2)}\,,\]
and
\[p_{l,d}\equiv\frac{l}{d-3}\,.\]
This is the simplified form of the master equation, from which all
metric components can be obtained.

The four regular singular points of the equation are:
$0$,$1$,$w_{l,d}$ and $\infty$. This form of Fuchsian equation is
known as Heun's equation (see appendix~\ref{ap A} for a review of
Fuchsian, Hypergeometric and Heun's equations including the Riemann
\emph{P}-Symbol). It can be characterized by the Riemann
\emph{P}-Symbol
\[
P \left (\begin{array}{ccccccccc}
0 \;& & 1 \;& & w_{l,d} \;& & \infty \;& &\\
0 \;& & 0 \;& & 0 \;& & -p_{l,d} \;& & ;\ X,\;\;\; q_{l,d} \\
-1 \;& & 0 \;& & 2 \;& & 1+p_{l,d} \;& &
\end{array}
\right ).\]
where $q_{l,d}\equiv-p_{l,d}\,(p_{l,d}+1)\,(d-4)\,w_{l,d}$.

The exponent difference is an integer number at all the points, except
for infinity. Thus, the solutions can be represented at each singular
point (except for infinity) in two forms of power series expansions:
one which is just a regular Taylor series (the exponent is zero), and
a second which may have a logarithmic divergence at the singular
point. At infinity, the difference of the exponents is an integer when
$l$ is a multiple of $\frac{d-3}{2}$. Then we have one solution
without a log singularity at infinity and second solution which may
have a log singularity at infinity. Otherwise, the two exponents
correspond to two different solutions without log at infinity. For
physical perturbations of the black hole we will consider only the
regular solutions at the horizon which correspond to the regular
solution at the singular point $X=1$.

\pagebreak[3]

The characteristic exponents in Heun's equation are determined by 4
parameters of the equation (see appendix~\ref{ap A}) which we denote
by $\alpha,\beta,\gamma,\delta$ . In our case
\begin{eqnarray*}
\alpha&=&-p_{l,d}\,,
\\
\beta&=&1+p_{l,d}\,,
\\
\gamma&=&2\,,
\\
\delta&=&1\,.
\end{eqnarray*}
Our case is a rather special case of Heun's equation because its
solutions have a very tight relation to the solutions of an
Hypergeometric equation with the same $\alpha,\beta,\gamma$ parameters
(the Hypergeometric equation is determined by 3 parameters, see
appendix~\ref{ap B}). The Riemann \emph{P}-Symbol for this
Hypergeometric equation is the following
\[ F_{l}(X)=
P \left (\begin{array}{ccccccc}
0 \;& & 1 \; & & \infty \;& &\\
0 \;& & 0  \;& & -p_{l,d} \;& & ;\ X \; \;\\
-1 \;& & 1 \;& & 1+p_{l,d} \;& &
\end{array}
\right ).\]
The Hypergeometric equation for $F_{l}(X)$ is
\begin{equation} \label{Hyper}
X\,(1-X)\,\frac{d^{2}\,F_{l}(X)}{dX^{2}}+2\,(1-X)\,\frac{d\,F_{l}(X)}{dX}+p_{l,d}\,(1+p_{l,d})=0\,.
\end{equation}

Let us define the linear operator $\mathcal{L}_{\rm Huen}^{l}$ as
\begin{equation} \label{defope}
\mathcal{L}_{\rm Huen}^{l}\equiv
\frac{d^{2}}{dX^{2}}+\left(\frac{2}{X}+\frac{1}{X-1}-\frac{1}{X-w_{l,d}}\right)\,\frac{d}{dX}-p_{l,d}\,(1+p_{l,d})\,\frac{X+(d-4)\,w_{l,d}}{X\,(X-1)\,(X-w_{l,d})}\,,
\end{equation}
so that~(\ref{keq}) can be written as
\[\mathcal{L}_{\rm Heu<n}^{l}\,E_{l}=0\,,\]
and the linear operator $\mathcal{L}_{\rm Hyper}^{l}$
\[\mathcal{L}_{\rm Hyper}^{l}\equiv X\,(1-X)\,\frac{d^{2}}{dX^{2}}+2\,(1-X)\,\frac{d}{dX}+p_{l,d}\,(1+p_{l,d})\,,\]
so that~(\ref{Hyper}) can be written as
\[\mathcal{L}_{\rm Hyper}^{l}\,F_{l}(X)=0\,.\]
By defining two additional first order linear operators
\begin{eqnarray}
\mathcal{S}_{1}&\equiv& 1-(d-3)\,X\,\frac{d}{dX}\,,
\\
\mathcal{S}_{2}^{l}&\equiv&
\frac{d-3}{X-1}\,\frac{d}{dX}-\frac{w_{l,d}\,(d-4)+X}{X\,(X-1)\,(X-w_{l,d})}\,,
\end{eqnarray}
one can verify by direct calculation that
\begin{equation}
\mathcal{L}_{\rm Heun}^{l}\,\mathcal{S}_{1}=\mathcal{S}_{2}^{l}\,
\mathcal{L}_{\rm Hyper}^{l}\,.
\end{equation}
Therefore, if $F_{l}(X)$ is a solution of the Hypergeometric
equation~(\ref{Hyper}), then $\mathcal{S}_{1}F_{l}(X)$ is a solution
of Heun's equation~(\ref{keq}). Namely, the solutions of~(\ref{keq})
can be written as
\begin{equation} \label{thesol}
E_{l}(X)=F_{l}(X)-(d-3)\,X\,\frac{d\,F_{l}(X)}{dX}\,.
\end{equation}
Notice that the singular point at $w_{l,d}$ does not exist in the
Hypergeometric function. It is added to the solution through the
operator $\mathcal{S}_{2}^{l}$.

The solution regular at the horizon $X=1$ written in terms of a
hypergeometric function is
\begin{equation} \fbox{$\room \displaystyle
E_{l} =\left(1-(d-3)\,X\,\frac{d}{dX}\right)(1-X)\,_{2}\!F_{1}
\left(1-\frac{l}{d-3},2+\frac{l}{d-3},2\,;1-X\right),$}
 \label{multfun} \end{equation}
and is our final expression for the solution of the master
equation~(\ref{geneq}). Note that when $\frac{l}{d-3}$ is a
natural number the solution for $E_{l}(X)$ is a polynomial and
thus regular at all 3 singular points. We add here the expression
for $A_{l}$~(\ref{Aneq}) written in terms of the variable~$X$:
\begin{equation} \fbox{$\room \displaystyle
-A_{l}=\frac{(d-3)^{2}\,(d-2)\,(X-1)}{2\,l\,(l+d-3)}
\left(2\,\frac{dE_{l}}{dX}
+X\frac{d^{2}E_{l}}{dX^{2}}\right) +\frac{d-4}{2}\,E_{l} \,.$}
\end{equation}

\section{The matching procedure}
\label{MatchingProcedure-section}

\subsection{Transformation of coordinates: Schwarzschild to harmonic} \label{trans}

The choice of harmonic coordinates in the asymptotic zone as opposed
to Schwarzschild coordinates in the near zone leads us to consider
carefully the transformation of coordinates in order to compare the
multipole moments in the overlap region correctly. The harmonic
condition~(\ref{gaugeE}) is satisfied in the asymptotic flat region
only in cartesian coordinates. Thus, we have to work in coordinates
which are cartesian in spatial infinity. Since the Schwarzschild
coordinates are spherical coordinates at infinity we take the $d-1$
cartesian coordinates that are related to
$(\rho,\chi,\theta_{1},\ldots ,\theta_{d-3})$ where $\rho$ is the
Schwarzschild radial coordinate.

Now one can verify that by the transformation of the form
$\rho\rightarrow\rho_{h}(\rho)$ we are able to transform the
Schwarzschild coordinates to harmonic coordinates, provided
\begin{equation}
\frac{d}{d\rho}\left(\left[\rho^{d-2}-\rho\,\rho_{0}^{d-3}\right]\,\frac{d\,\rho_{h}}{d\rho}\right)=(d-2)\,\rho^{d-4}\,\rho_{h}\,.
\label{gauged}
\end{equation}
(\ref{gauged}) is obtained by substituting any of the expressions for
$x^{\mu}$ expressed in spherical coordinates, where we take
$\rho_{h}(\rho)$ instead of $\rho$, into the definition of the
harmonic gauge (static case)
\begin{equation}
\triangle x^{\nu}= 0\,.
\end{equation}
Without loss of generality we take
\[\triangle z=\triangle \left( \rho_{h}(\rho)\,\cos(\chi) \right)= 0\,.\]
Since the laplacian of any of the cartesian coordinates reduce
to~(\ref{gauged}), we know that the transformation
$\rho\rightarrow\rho_{h}(\rho)$, when $\rho_{h}(\rho)$ is defined as a
solution of~(\ref{gauged}), gives the required
transformation. Equation~(\ref{gauged}) has two solutions which we can
write as a power series. When $\rho \gg \rho_{0}$,~(\ref{gauged})
becomes
\[\frac{d}{d\rho}\left(\rho^{d-2}\,\frac{d\,\rho_{h}}{d\rho}\right)=\rho_{h}\,\rho^{d-4}\,(d-2)\,.\]
Substituting $\rho_{h}=\rho^{k}$ we get a quadratic equation for the
characteristic exponents at infinity (see appendix~(\ref{ap A})~)
\[k^{2}+k\,(d-3)-(d-2)=0\,,\]
whose solutions are
\[k=1, -(d-2)\,. \]
We set one boundary condition that for large $\rho$, $\rho_{h} \sim
\rho$, i.e., for large $\rho$ we expect that the two gauges
coincide. The leading terms in the transformation are
\begin{equation} \label{gauge_ex}
\rho_{h}=\rho-\frac{\rho_{0}^{d-3}}{2\,(d-3)\,\rho^{d-4}}+\mathcal{O}\left(\frac{1}{\rho^{d-3}}\right).
\end{equation}

\subsection{Zones and basic dialogue arrows}
\label{desc}

A black hole in a spacetime with compact dimension is distorted due to
tidal fields that originate from its effective mirror images. Namely,
the periodic boundary conditions in the compact dimension change the
structure of the solution compared to the case of flat asymptotic
infinity. We assume that the black hole is small compared to the
period of the compact dimension $L$ ($\rho_{0} \ll L$). Hence in the
limit $\rho\sim\rho_{0} \ll L$ the exact solution should become a
$d$-dimensional Schwarzschild solution. For $\rho_{0}\ll \rho\sim L$,
far from the black hole, we have a flat $d$-dimensional Minkowsky
spacetime with a periodic coordinate (which we denote by $z$).

In the intermediate region we can approximate the exact solution by
matching two perturbative expansions from both sides. This procedure
can be viewed as a dialogue of multipoles where the black hole changes
its shape (mass multipoles) in response to the field (multipoles)
created by its periodic ``mirrors'', and that in turn changes its
field and so on. Usually when we use an iterative method for solving
non-linear equations, we have to supply enough data about the boundary
conditions at each step of the iteration so that the solution would be
determined completely ,i.e., without any free parameters or
functions. Here the two known solutions (the Schwarzschild solution
and the Minkowsky spacetime with a periodic coordinate) will play the
role of boundary conditions for our problem. Hence, each perturbative
expansion will complete the missing information in the other
expansion, order by order. After describing the method we will show
that it is well-posed, namely that all the information needed for a
certain perturbation order is provided by the previous ones.  The
approximate solution in the end will be written as a combination of a
post-newtonian expansion far away from the black hole and an expansion
of the same order around the asymptotically flat black hole
solution. For this purpose, we first divide the spacetime into two
zones with an overlap region (see figure~\ref{areas}):

{\renewcommand\belowcaptionskip{1em}
\FIGURE[t]{\epsfig{file=Areas.eps, width=.7\textwidth, clip=}%
\caption{The division of the spacetime into two overlapping zones: the
  \emph{near zone} $\rho \gtrsim \rho_{0}$ where the perturbative
  parameter is $L^{-1}$, the \emph{asymptotic zone} $\rho_{0}\ll
  \rho \sim L$ where the perturbative parameter is $\rho_{0}$ and the
  \emph{overlap region} in between.\label{areas}}}}

\paragraph{The near zone:} for $\rho \gtrsim \rho_{0}$ we expand
the solution around the $d$-dimensional Schwarz\-schild solution
characterized by $\rho_0$ (see section~\ref{BHperturb}). Since
$\rho\sim\rho_{0}$ in this zone, the small dimensionful parameter of
the expansion in this zone would be $L^{-1}$. Notice that when
$L\rightarrow \infty$ (i.e., the limit of ``unfolding" the compact
dimension) we are left only with the first term (zeroth order) of the
expansion which is the $d$-dimensional Schwarzschild solution.  The
next terms represent the influence of the compact dimension on the
black hole's metric. From the point of view of the near zone, as the
period $L$ is large compared to $\rho_{0}$, the terms in the expansion
series become smaller and the convergence of the perturbation series
is better.

\paragraph{The asymptotic zone:} for $\rho_{0}\ll \rho \sim L$ we
expand around the flat background with a compact dimension of size
$L$. Here we take $\rho_{0}$ the size of the black hole as the small
dimensionful parameter (this is the so called post-newtonian
expansion, see section~\ref{newtoniansec}, with the matching added as
a boundary condition).  The terms in the expansion represent the
distortion of the flat spacetime (with a compact dimension) due to the
presence of the black hole and its ``mirror images". From the point of
view of the asymptotic zone, as the mass of the black hole becomes
smaller, the terms in the expansion series become smaller and the
convergence of the perturbation series is better.

\paragraph{The overlap region:} each one of the expansions cannot
be determined independently since it requires boundary conditions
at the overlap. Hence, it carries free parameters that should be
determined by matching to the other expansion in the overlap
region $\rho_{0}\ll \rho \ll L$. This is achieved by comparing the
multipoles of the expansions order by order; In the asymptotic
zone we look at the multipoles in the limit $\rho \ll L$ and match
them with the multipoles of the near zone expansion in the limit
$\rho \gg \rho_{0}$.

\pagebreak[3]

There are three parameters that determine the matching procedure for
each term in the multipole expansion:
\begin{itemize}
\item The order of the multipole expansion which we denote by $n$. A
  term of order $n$ in the asymptotic zone expansion is proportional
  to $\rho_{0}^{n}$ while a term of the same order in the near zone is
  proportional to $({1}/{L})^{n}$.
\item The multipole number $l$. In each order a given multipole can
  have contributions from lower multipoles (via non-linear sources
  from the same zone, e.g.\ addition of angular momenta) and from
  matching which is linear and as such can be performed only between
  identical multipoles in both zones.
\item The number of dimensions $d$.
\end{itemize}
The interplay between these three parameters $l,n,d$ creates the
matching pattern which we describe below.

In the previous sections we obtained the linear perturbations around
the exact solutions (equations~(\ref{keq})
and~(\ref{Newtonian-metric})). In order to get non-linear corrections
by the method of successive approximations we substitute in Einstein's
equations expansions of the form
\begin{equation} \label{metricexp}
g_{\mu\nu}=g^{(0)}_{\mu\nu}+g^{(1)}_{\mu\nu}+g^{(2)}_{\mu\nu}
+\cdots +g^{(n)}_{\mu\nu}+\cdots \,,
\end{equation}
where $g^{(0)}_{\mu\nu}$ is the exact solution.  For each term of this
expansion we get an equation for the metric components of the form
\begin{equation} \label{opeLin}
\mathcal{L}(g^{(n)}_{\mu\nu})=\mathcal{F}_{n}(g^{(n-1)},g^{(n-2)},
\ldots ,g^{(1)})\,,
\end{equation}
where the superscript $n$ denotes the order in the perturbation,
$\mathcal{L}$ is linearized operator (which is second order and
independent of $n$) and $\mathcal{F}_{n}$ stands for nonlinear terms
that are functions of the previous orders and their derivatives. Since
the asymptotic zone expansion is in the harmonic gauge, $\mathcal{L}$
there is simply the laplacian. In the near zone, as we saw in
section~\ref{BHperturb}, the operator is more involved and it is given
by $\mathcal{L}^{l}_{\rm Heun}$, defined in~(\ref{defope}). At each
order one is free to add to the metric a solution to the homogeneous
equation (an element of $\mbox{ker}(\mathcal{L})$). The homogeneous
solution of any order $n$ can be expanded into multipoles. Thus we can
write the general form of the $n$-th correction to the metric in the
form
\begin{equation}
\label{expanmulti}
g_{\mu\nu}^{(n)}=g_{\mu\nu,(p)}^{(n)}+\sum_{l=0}^{\infty}
c^{(n)}_l\, f^{(n)\,l}_{\mu \nu}(\rho)\,\Pi_{0}^{l,d}(\chi)\,.
\end{equation}
The first term is the particular solution of order $n$. The second
term is the multipole expansion of the homogeneous solution where
$\Pi_{0}^{l,d}(\chi)$ is the spherical harmonics function, defined
in~(\ref{pieq}) (remember that we have $\SO(d-2)$ symmetry).
$f^{(n)\,l}_{\mu \nu}(\rho)$ is the multipole radial function and
$c^{(n)}_l$ are coefficients that we should determine using the
matching procedure for each multipole $l$ and order $n$ (in $d$
dimensions). For the matching we would have to expand the particular
solutions to multipoles as well.

>From the previous subsection we learn that in the limit $\rho \gg
\rho_{0}$ the Schwarzschild and harmonic coordinates coincide. Hence
we are allowed to compare leading terms in the two zones without
coordinate transformations. Moreover, in this limit ($\rho \gg
\rho_{0}$) the homogeneous linear operator in the near zone becomes
the laplacian (Einstein's equations in this limit coincides with the
newtonian poisson equations). Therefore, in both zones, the leading
terms will be multipole functions of the laplacian, namely, leading
terms that their dependence on $\rho$ goes like $\rho^{l}$ or
$\rho^{-(l+d-3)}$ (see~(\ref{multilap})).

In the homogeneous solutions of the near zone the leading terms (the
ones to be matched) behave like $\rho^{l}$ (when one takes the limit
$\rho \gg \rho_{0}$). The component with the subleading behavior
$\rho^{-(l+d-3)}$ is determined by the boundary condition of
regularity at the horizon. This can be seen explicitly from the
solution that we have for $E_{l}$ in section~\ref{solution}. The
Riemann \emph{P}-Symbol (appendix~\ref{ap A}) shows us that the
asymptotic behavior of the two power series expansions\footnote{Here
  for simplicity of the matching procedure we consider the
  equation~\ref{keq} with dimensions, namely, $X \rightarrow X \cdot
  \rho_{0}^{d-3}$.} at infinity is $\rho^{l}$ or
$\rho^{-(l+d-3)}$. For the asymptotic zone the role is reversed -- the
leading term in the overlap zone ($\rho \ll L$) is $\rho^{-(l+d-3)}$
while the subleading term $\rho^{l}$ is fixed by the boundary
conditions at infinity.

We summarize the discussion above using a graphic representation of
the multipole expansion for multipole number $l$ in
figure~\ref{config}. The numbers stand for the order in the multipole
expansion. The two rows of cells represent the expansion in the
asymptotic zone (above) and in the near zone (below). The left side of
each cell contains the leading term in the overlap region (that should
``receive" the information from the matching). The right side of each
cell contains the appropriate term to match the left side of different
cell. The horizontal part in the cell reminds us the order of
perturbation --- $\rho_{0}^{n}$ in the asymptotic zone and $L^{-n}$ in
the near zone. The matching procedure actually becomes a process of
determining which cell's left side receives information from which
cell's right side (as we will see, not all the cells are relevant for
this process).

\FIGURE[t]{\epsfig{file=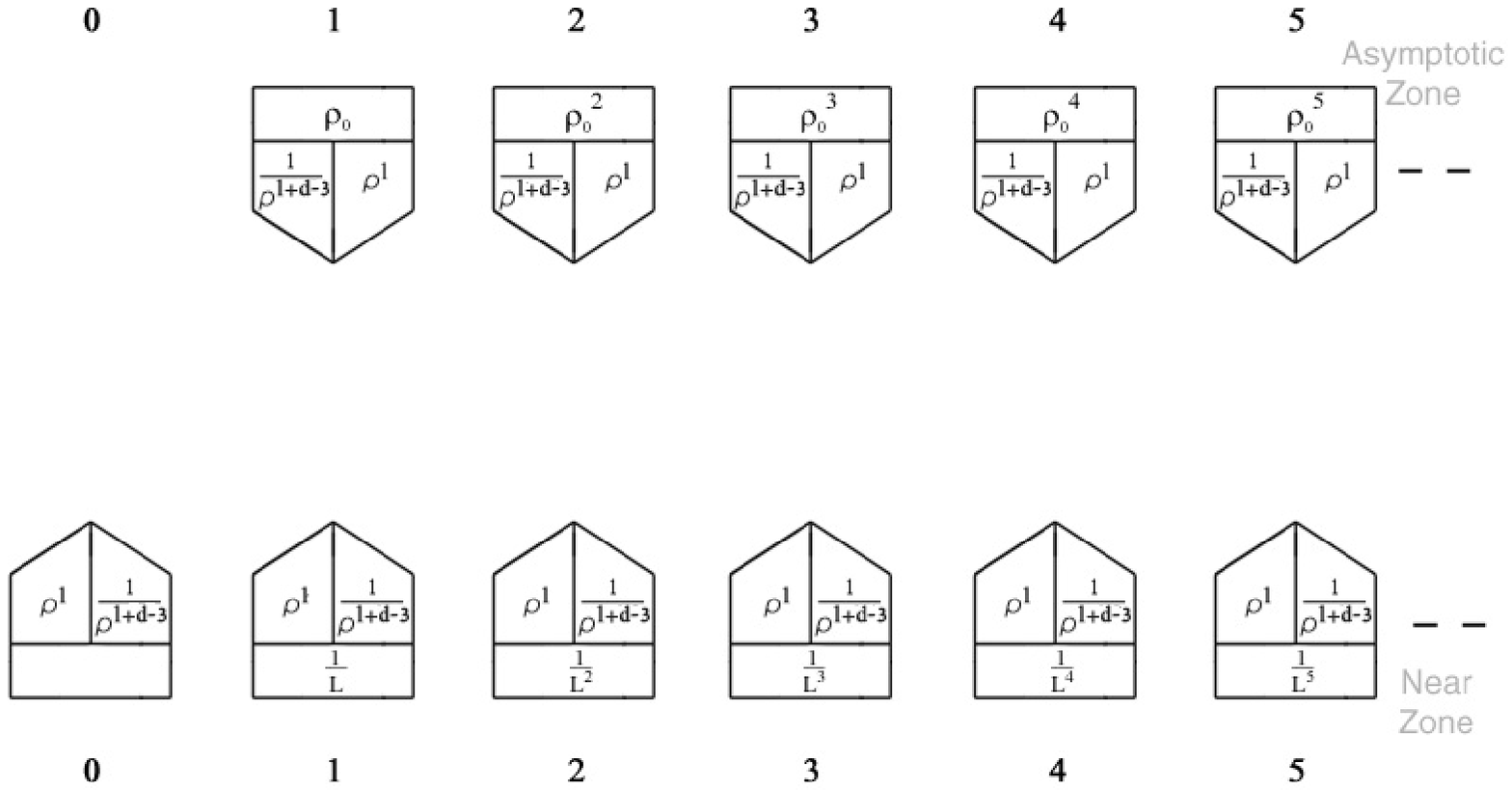, width=\textwidth, clip=}%
\caption{The representation of the expansion in the near and
  asymptotic zones for a multipole moment $l$, before the
  matching. Each cell in the upper row stands for a term in the
  asymptotic zone expansion and each cell in the lower row stands for
  a term in the near zone expansion. Each cell consists of the small
  parameter in the appropriate power and two terms to be
  matched.\label{config}}}

To determine the pattern of perturbation, namely the basic ``dialogue
arrows'' of information flow in a diagram such as figure~\ref{config},
the basic observation comes from dimensional analysis. A mass
multipole moment of multipole number $l$, $\mathcal{M}_{l}$, is
defined as
\[\mathcal{M}^0_{l_{~i_1,\dots,i_n}} \sim \int\! dm \,
x_{i_{1}}x_{i_{2}}\ldots {x_{i_{l}}}\,,\] where the indices $i_{j}$
denote the cartesian coordinates, and the superscript $0$ is a
reminder that these are multipoles at the origin. The gravitational
field of multipole of order $l$ of mass distribution at the origin
behaves like
\[\frac{\mathcal{M}^0_{l}}{\rho^{l+d-3}}\,, \]
and the multipole moment itself has the length dimensions
\begin{equation}
\left[\mathcal{M}^0_{l}\right]=l+d-3\,. \label{MultDim0}
\end{equation}
Such mass multipoles at the origin are used to supply matching
boundary conditions for the asymptotic zone. Suppose we are at order
$n$ in the asymptotic zone so that
\begin{equation}
\mathcal{M}^0_{l}  \sim {\rho_0^{~n} \over L^{m}}\,,
\label{MultOrder}
\end{equation}
where $m$ is the near zone order from which it was
matched. Comparing~(\ref{MultDim0}) and~(\ref{MultOrder}) we conclude
that
\begin{equation}
 n-m= \left[\mathcal{M}^0_{l}\right]=l+d-3\,,
\end{equation}
namely, upward pointing ``dialogue arrows'' have a step-size of
$l+d-3$.  In a similar manner the multipole moments at infinity (used
to supply boundary conditions for the near zone) yield
\[ n-m=\left[\mathcal{M}^{\infty}_{l}\right]= -l\,,\]
since the (dimensionless) gravitational field behaves like
\[\mathcal{M}^{\infty}_{l}\rho^{l}\,.\]
Namely, downward pointing ``dialogue arrows'' have a step-size of
$+l$.

Alternatively, the same result can be described from the following
point of view. In the perturbation process we expand a dimensionless
quantity which satisfied Laplace's equation in the two zones and
compare the expansions in the overlap zone. In the matching process we
identify two terms that make one dimensionless quantity in the overlap
region. The process is demonstrated in figure~\ref{dia2}. Looking at a
cell of order $m$ in the near zone, we see that the corresponding term
in the expansion is proportional to
$\frac{1}{L^{m}\,\rho^{l+d-3}}$. In order to create a dimensionless
quantity, we have to match this term to order $n=l+m+d-3$ in the
asymptotic zone. Then we have a leading term proportional to
\[\frac{\rho_{0}^{l+m+d-3}}{L^{m}\,\rho^{l+d-3}}\,,\]
up to a numerical constant. Using the same argument, this cell in the
asymptotic zone is matched to the cell $m'=n+l=m+2l+d-3$ in order to
create the dimensionless quantity
\[\frac{\rho_{0}^{n}\,\rho^{l}}{L^{n+l}}\,.\]

\FIGURE[t]{\epsfig{file=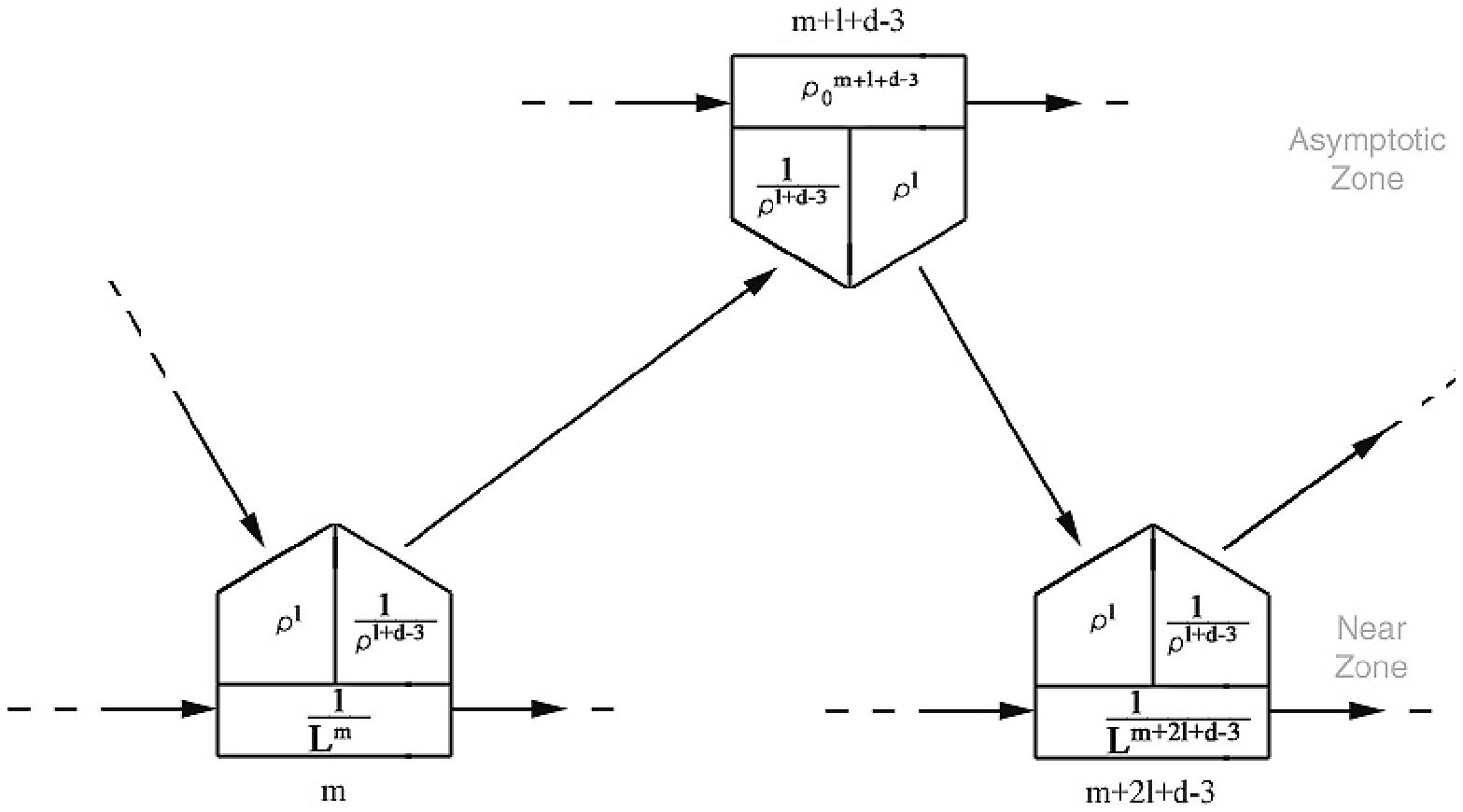, width=.9\textwidth, clip=}%
\caption{Two steps in the matching procedure. A term of order $m$ and
  multipole number $l$ in the near zone determines the multipole of
  order $n=m+l+d-3$ in the asymptotic zone. Afterwards the latter is
  used to determine a term of order $m'=n+l=m+2l+d-3$ in the near
  zone.\label{dia2}}}

Apparently, this process can continue to create an infinite chain of
matchings for each multipole which consists of two unequal steps; from
the near to asymptotic zone we have a step of $\Delta n=l+d-3$ orders
and from the asymptotic to the near we have a step of $\Delta n =l$
orders.

\subsection{Pattern of the dialogue}

Let us look in detail into the pattern of the dialogue. We would like
to summarize all the ``information flow arrows'' between the various
orders and to conclude how the computation should be performed to
determine the metric up to a prescribed order in one of the zones
(following what we term ``the critical route'').

First we note that since our problem is defined with reflection
symmetry $z\rightarrow -z$, one should consider only multipoles that
satisfy this symmetry. Applying the transformation $\chi \rightarrow
\chi + \pi$ in~(\ref{rodrig}) shows that \emph{only even multipoles}
are relevant for our expansions.

We summarize the information flow arrows for a diagram such as
figure~\ref{config}:
\begin{itemize}
\item Within a zone.

These are the ordinary non-linear sources of the perturbation
procedure. At order $n$ there can be a non-zero source only if there
are some non-zero metrics at orders $n_i<n$ such that some linear
combination of $n_i$ with integer positive coefficients yields~$n$.

\item Inter-zone arrows.

We saw above that arrows going ``down'' (from the asymptotic to the
near zone) advance by $l$ orders, while those going in the reverse
direction advance by $l+d-3$ orders. Since $l$ can only be even there
are \emph{two basic steps: $2$ and $d-3$}.
\end{itemize}

Finally there is an ``initial condition'': the leading correction to
the zeroth order metrics is the newtonian potential at order $d-3$ in
the asymptotic zone. Therefore the larger $d$ is there will be more
orders where the metric vanishes.

>From a practical point of view it is especially important to determine
the ``critical route''. Namely, suppose we wish to compute the metric
up to a prescribed order $n$ in a certain zone.  Which orders must be
determined on the way? Clearly we should know all the lower orders in
that zone.\footnote{Actually, this may not be necessary since for
  example orders $n'$ such that $n-(d-3)<n'<n$ will not
  contribute. Yet, normally, we are not interested in a correction
  before all previous ones were determined.} In addition we should
have information from the other zone. If the original zone in which we
were interested is the near zone we should know the metric at the
asymptotic zone at orders $n-l$ for all even $l$, and similarly if it
were the asymptotic one we need to know all the $n-(d-3)-l$ orders of
the near zone. We see that \emph{in both cases the limiting $l$ is the
  monopole} $l=0$ which sets the critical route to arrive from the
other zone: \emph{to reach the near (asymptotic) order $n$ the
  critical route passes through the asymptotic (near) order $n$
  ($n-d+3$)}. It is clear that the procedure does not run into closed
loops since the sum of the two step-sizes, being $d-3$, is greater
than 0, and in that sense \emph{there are sufficient boundary
  conditions to determine the expansion, namely it is well-posed.}

In order to make this analysis more concrete we shall first outline
the first steps in the procedure for $d=5$ which is somewhat special
and then do the same for some general $d>5$.

In the case of 5 dimensions $d-3=2$ and this implies that only the
even orders are involved in the expansion (the same holds true for any
odd dimension). Hence in 5 dimensions we actually expand in the small
parameters $\rho_{0}^{2}$ and $\frac{1}{L^{2}}$ (see
figure~\ref{5dialogue}). This is a great simplification to the
perturbation scheme. Moreover, all the even multipole functions
(see~(\ref{multfun})) are polynomials in 5 dimensions.

In figure~\ref{5dialogue} we show the first steps in the matching
procedure in 5 dimensions, until the fourth order.  In order to have
the metric until the fourth order in the two zones we need, besides
the particular solutions, to match the monopole ($l=0$) and the
quadrupole ($l=2$). In order to obtain higher orders we will have to
add higher multipoles to the matching. Note that we have to know the
fourth order in the asymptotic zone in order to get the fourth order
in the near zone. This situation is rather different for $d>5$.

\FIGURE[t]{\epsfig{file=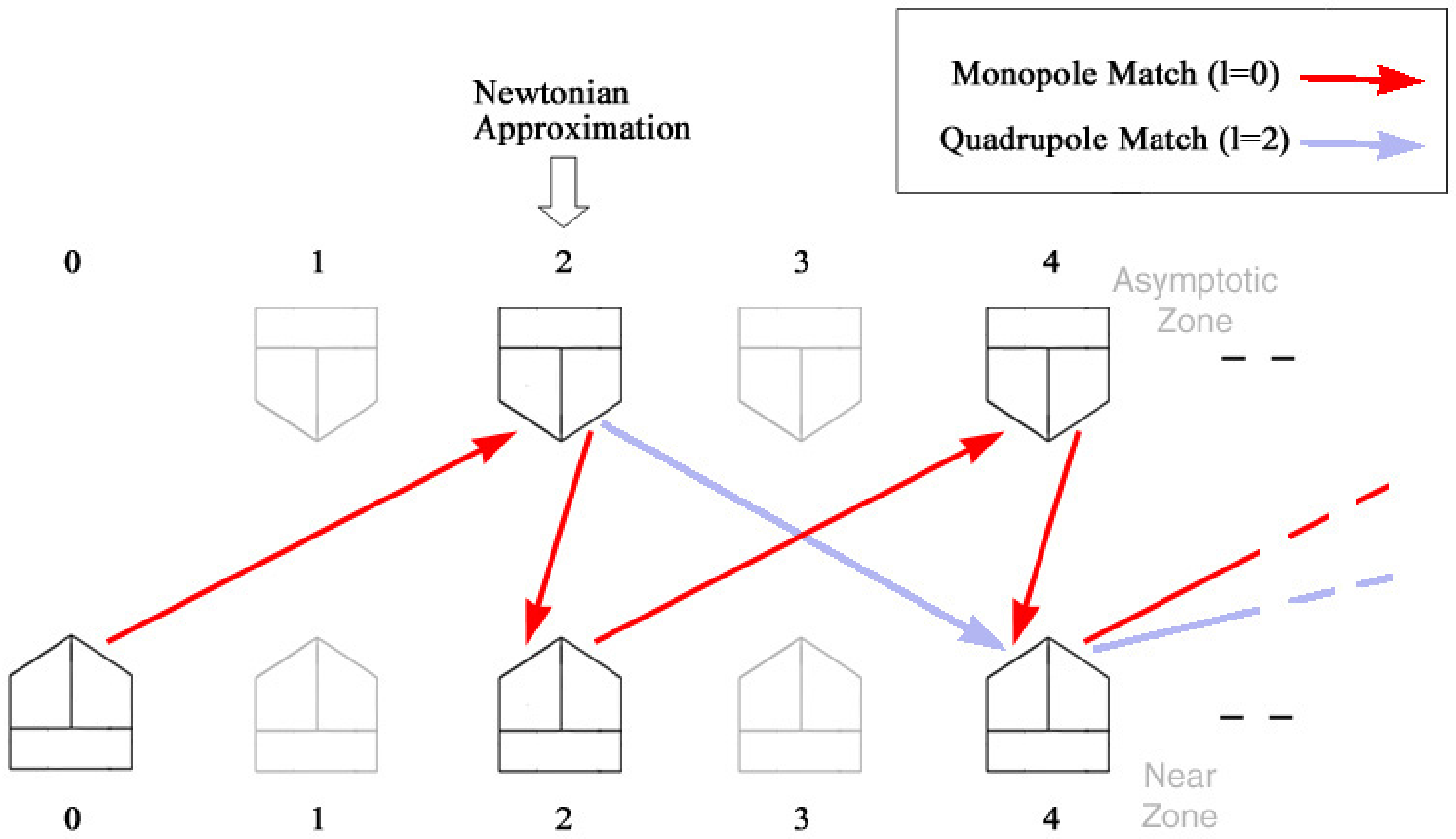, width=\textwidth, clip=}%
\caption{The first steps in the matching procedure in 5
  dimensions. Until the fourth order only monopole and quadrupole
  matches appear in the two zones. Each one of the matches is
  represented by a different color. The newtonian approximation enters
  in the second order in the asymptotic zone. It is clear from the
  figure that the monopole is the limiting and hence the most ``fresh"
  multipole that should be determined for a certain
  order.\label{5dialogue}}}

In 5d it is easy to understand that the method is well-posed. If we
look at a cell of order $n$ in the near zone, then all the (even)
multipoles from $l=0$ to $l=n$ are determined from the previous orders
in the asymptotic zone when the monopole is the last to be matched ---
it is determined from the same order but in the asymptotic zone. We do
not have to determine any multipole higher than $n$ because they do
not appear at this order --- in order to create a dimensionless
quantity out of $\rho^{n+k}$ for $k\geq1$ we have to multiply it at
least by $\frac{1}{L^{n+k}}$ (thus this multipole will appear at order
$n+k$ at least in the near zone). Therefore, for any order until $n$
the metric in the near zone is determined. Similar argument is valid
for the asymptotic zone as well. let us look at a cell of order $n$ in
the asymptotic zone. All the multipoles from $l=0$ to $l=n-2$ are
determined by matching to the near zone when $l=0$ (the monopole) is
the last one to be matched (it is matched with the cell of $2$ orders
before) --- see figure~\ref{5dialogue}. Therefore, the information for
any cell comes from the cells of lower orders in the other zone in 5d.

{\renewcommand\belowcaptionskip{-1em}
\FIGURE[t]{\epsfig{file=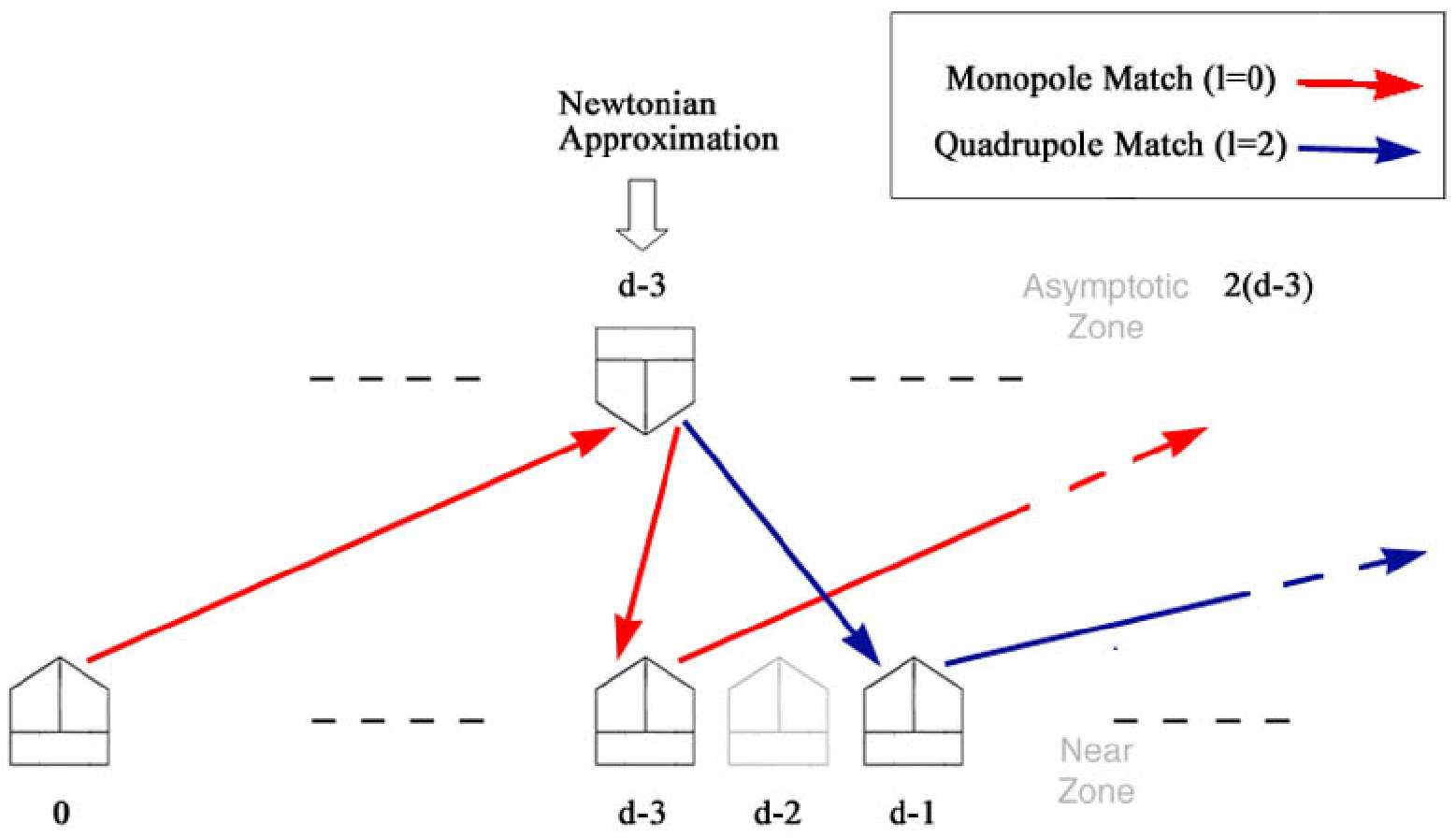, width=\textwidth, clip=}%
\caption{The first steps in the matching procedure for $d>5$. Compare
  with the previous figure (the same in 5 dimensions). Here the first
  quadrupole matching in the near zone is done before the second
  correction to the monopole.\label{ddialogue}}}}

The first steps in the matching for $d>5$ dimensions are shown in
figure~\ref{ddialogue}. Since the newtonian approximation enters at
the order $d-3>2$, the post-newtonian correction comes at
$2\,(d-3)$. The first quadrupole contribution in the near zone comes
in $(d-3)+2=d-1$. Then, as opposed to the case of $d=5$, the second
matching of the monopole in the near zone comes after the quadrupole
matching, since $2\,(d-3)>d-1$. As we go to higher dimensions, there
are more multipoles in the near zone that we can match with the
multipoles in the newtonian approximation before adding the second
correction to the monopole. This can be calculated precisely: The
length of a step of a match from the asymptotic zone to the near zone
is $\Delta n=l$. The distance between the newtonian and post-newtonian
approximations is $\Delta n=d-3$. So any multipole that satisfy
$l<d-3$ comes before the second correction to the monopole. An
interesting limit is when $d \rightarrow \infty$. Then, the newtonian
approximation is enough to determine all the orders of expansion in
the near zone.

\section{Matching results}
\label{results}

After describing the procedure in general, let us start and calculate
the first coefficients of the approximate solution. Note that in the
matching process it is enough to match a single metric component to
guarantee the matching of the whole metric. For our purposes it will
be convenient to consider for the matching the dimensionless quantity
${g_{\chi\chi}}/{\rho^{2}}$.

\subsection{The first monopole match}

Actually there is a zeroth monopole match which relates the mass
measured asymptotically (through the newtonian potential) with the
$\rho_0$ parameter of the \Schw\ solution, namely a matching between
the zeroth order in the near zone and order $n=d-3$ in the asymptotic
zone (the first red arrow in figures~\ref{5dialogue}
and~\ref{ddialogue}). This match was done in
section~\ref{newtoniansec} (based on~\cite{myers}) and gives the
identification $\rho_{0}^{d-3}=\frac{16 \,\pi
  \,G_{d}\,M}{(d-2)\,\Omega_{d-2}}$.

The first monopole match is between the monopole in the newtonian
approximation ($n=d-3$) and the first monopole correction to the
$d$-dimensional Schwarzschild solution. Being the $l=0$ correction the
spherical symmetry of the black hole will not be lost, but rather the
physical interpretation is that the black hole is placed in a region
with a constant non-zero gravitational potential due to the array of
image black holes.

First we expand the newtonian approximation\footnote{Recall that
  $\rho$ in the asymptotic zone is in harmonic gauge while $\rho$ in
  the near zone is in the ``no derivative'' gauge around the
  Schwarzschild gauge.  Since we compare only leading terms in the two
  zones and in order to simplify the notation we denote both of them
  by $\rho$.} in multipoles around the origin including not only the
monopole term but also the quadruple in anticipation of the
computation of eccentricity. The expansion around $\rho=0$ in polar
coordinates $(\rho,\chi)$ is
\begin{eqnarray}
\frac{g_{\chi\chi}^{\mbox{(asymp)},d-3}}{\rho^{2}}&=&
\frac{\rho_{0}^{d-3}}{d-3}\sum_{n=-\infty}^{\infty}\frac{1}{(r^{2}
+(z-nL)^{2})^{\frac{d-3}{2}}}
\nonumber\\
&=&\frac{1}{d-3}\cdot\frac{\rho_{0}^{d-3}}{\rho^{d-3}}+\frac{2}{d-3}\cdot
\frac{\rho_{0}^{d-3}}{L^{d-3}}\,\zeta(d-3)+
\nonumber\\
&& +\frac{(d-2)\,\rho^{2}\,\rho_{0}^{d-3}}{4\,
L^{d-1}}\,\zeta(d-1)\,\Pi^{2,d}_{0}(\chi)+\mathcal{O}(\rho^{4})\,,
\label{mono2}
\end{eqnarray}
where $\zeta$ is Riemann's zeta function\footnote{Riemann's zeta
  function is defined as
  $\zeta(s)=\sum_{n=1}^{\infty}\frac{1}{n^{s}}$.} and
\[\Pi^{2,d}_{0}(\chi)=\frac{1}{d-2}\,\left(\cos^{2}(\chi)\,(d-1)-1\right),\]
according to the Rodrigues formula~(\ref{rodrig}). The superscript
``$\mbox{(asymp)},d-3$'' indicates that this is the correction of
order $d-3$ in the asymptotic zone. In the same way we will denote the
expansions in the near zone by the superscript ``(near)''.  Note that
there is no dipole term in the expansion (or any odd multipole) due to
the symmetry $\chi \rightarrow \chi+\pi$.

Recall the expressions that we had for the metric monopole
perturbations~(\ref{allo1})--(\ref{allo3}). Comparison with the
monopole term in~(\ref{mono2}) gives us the value of the matching
constant $C_{1}$ for the first correction to the metric in the near
zone ($n=d-3$)
\[C_{1}=\zeta(d-3)\,\frac{\rho_{0}^{d-3}}{L^{d-3}}\,.\]

The final form of the metric in the near zone with the first monopole
correction is
\begin{equation} \fbox{$
\begin{array}{rcl}
\room  g_{\chi\chi,d}^{\mbox{near}}&=&\displaystyle
\rho^{2}\left(1+\frac{2\,\zeta(d-3)\,\rho_{0}^{d-3}}{(d-3)\,L^{d-3}}\right)
+\mathcal{O}\left(\frac{1}{L^{d-2}}\right),
\\
\room g_{tt,d}^{\mbox{near}}&=&\displaystyle
-\left(1-\frac{\rho_{0}^{d-3}}{\rho^{d-3}}\right)
\left(1-\frac{2\,\zeta(d-3)\,\rho_{0}^{d-3}}{L^{d-3}}\right)
+\mathcal{O}\left(\frac{1}{L^{d-2}}\right), \label{final_mono}
\\
\room g_{\rho\rho,d}^{\mbox{near}}&=&\displaystyle
\left(1-\frac{\rho_{0}^{d-3}}{\rho^{d-3}}\right)^{-1}
 \left( 1+\frac{2\,\zeta(d-3)\,\rho_{0}^{d-3}}{(d-3)\,L^{d-3}}
 \right) +\mathcal{O}\left(\frac{1}{L^{d-2}}\right).
\end{array}$}
\end{equation}

As mentioned above this monopole correction can be attributed to the
non-zero newtonian potential at infinity of the near zone. The same
result can be obtained in a less formal way by rescaling of $t$
reflecting in this way the change in the asymptotic boundary
conditions.

Since there is no change in the shape of the black hole we can measure
the effect of the correction by calculating the correction to the
dimensionless quantity $\kappa^{d-2}\,A_{d-2}$ where $\kappa$ is the
surface gravity of the horizon and $A_{d-2}$ is its
$(d-2)$-dimensional area.

\pagebreak[3]

To find the area we note that the horizon in this approximation is
located at a constant $\rho_{H}$ which is defined as the root of
$g_{tt}(\rho)$=0. Using the approximated metric above we get that the
horizon does not move (in our gauge)
\[
\rho_{H}^{d-3}=\rho_{0}^{d-3}+\mathcal{O}\left(\frac{1}{L^{d-1}}\right).
\]
The area of the horizon is
\begin{eqnarray}
A_{d-2}&=&g_{\chi\chi}(\rho_{H})^{\frac{d-2}{2}} ~ \Omega_{d-2}
\\
&=&\rho_{0}^{d-2}\,\Omega_{d-2}
\left(1+\frac{(d-2)\,\zeta(d-3)\,\rho_{0}^{d-3}}{(d-3)\,L^{d-3}}
+\mathcal{O}\left(\frac{1}{L^{d-1}}\right)\right).
\end{eqnarray}
The surface gravity calculated using~(\ref{kap}) of appendix~\ref{ap
  B} is
\[
\kappa=\left.\frac{1}{2}\,\frac{\mid \!\partial_{\rho}g_{tt}(\rho) \!\mid} {\sqrt{-g_{\rho\rho}(\rho)\,{g_{tt}(\rho)}}}\,\right|_{\rho=\rho_{H}}.\]
Substituting the approximate metric we get
\[\kappa=\frac{d-3}{2\rho_{0}}\,\left(1-\frac{(d-2)\,\zeta(d-3)}{d-3}\cdot\frac{\rho_{0}^{d-3}}{L^{d-3}}\right)+ \mathcal{O}\left(\frac{1}{L^{d-1}}\right).\]
Combining the results together we obtain
\begin{equation}
\fbox{$\room \displaystyle
\kappa^{d-2}\,A_{d-2}=\left(\frac{d-3}{2}\right)^{d-2}\,\Omega_{d-2}\,\left(1-(d-2)\,\zeta(d-3)\cdot\frac{\rho_{0}^{d-3}}{L^{d-3}}\right)+
\mathcal{O}\left(\frac{\rho_{0}^{d-1}}{L^{d-1}}\right).$}
\label{Akappa}
\end{equation}

\subsection{Matching multipoles to the newtonian approximation}
\label{MatchingMultipoles}

Our next goal is to obtain an expression for the deviation from
spherical symmetry --- the eccentricity of the horizon. For this
purpose we have to match the quadrupole term in the near zone to the
newtonian approximation. In this subsection we give a more general
recipe, matching any multipole to the newtonian approximation.

As was explained in the previous section, this matching is important
for higher multipoles as we increase the number of dimensions. From
the matching scheme we learn that the multipoles in the newtonian
approximation match with $\Delta n=l$ orders higher in the near zone
expansion.

Let us count the number of constants to match. Given $l$ there are two
radial solutions in the near zone. Regularity at the horizon fixes a
particular combination and so it remains to set the normalization of
this combination by comparing the leading terms from both zones. Thus
we need to determine a single constant per multipole.

In the near zone we have the multipole functions written in terms of
solutions of a Hypergeometric equation (recall~(\ref{Hyper})
and~(\ref{thesol})). Regularity of the metric at the horizon requires
that the relevant solution of the Hypergeometric equation be regular
there. The horizon is located at $X=1$ when we use the variable $X$ as
we defined in~(\ref{theX})
\[X=\left(\frac{\rho}{\rho_{0}}\right)^{d-3}.\]
Using the Riemann \emph{P}-Symbol of~(\ref{Hyper}) we can obtain the
regular solution at $X=1$ (see~\cite{erdel1}) and substitute it
into~(\ref{thesol})
\begin{equation}
E_{l}=\left(1-(d-3)\,X\,\frac{d}{dX}\right)\left(1-X)\,_{2}\!F_{1}(1-\frac{l}{d-3},2+\frac{l}{d-3},2\,;1-X\right).
\label{thetil}
\end{equation}
The correction to the near zone metric of order $l+d-3$ is
proportional to $E_{l}$
\begin{equation} \label{correction}
\frac{g_{\chi\chi}^{\mbox{(near)},l+d-3}}{\rho^{2}}=D_{l}\,\rho_{0}^{l}\,E_{l}(X)\,\Pi^{l,d}_{0}(\chi)\,,
\end{equation}
where $D_{l}$ is the matching constant which depends on the multipole
and the dimension. This constant will be determined by matching to the
asymptotic zone. If $d-3>l$ the newtonian approximation suffices and
otherwise non-linear terms from lower orders should be added.

In order to match this solution in the overlap zone we have to find
its leading term as $X \to \infty$. The regular solution of the
Hypergeometric equation~(\ref{Hyper}) at $X=1$ can be written as a
combination of the two solutions at infinity, using the identity
(see~\cite{erdel1})
\begin{eqnarray}
_{2}\!F_{1}(\alpha,\beta,\gamma;1-X)&=&c_{1}\,(X-1)^{-\alpha}\, _{2}\!F_{1}
\left(\alpha,\alpha-\gamma+1,\alpha-\beta+1;\frac{1}{1-X}\right)+
\nonumber\\
&&+c_{2}\,(X-1)^{-\beta}_{2}\!F_{1}
\left(\beta,\beta-\gamma+1,\beta-\alpha+1;\frac{1}{1-X}\right),
\end{eqnarray}
where
\begin{eqnarray*}
c_{1}&=&\frac{\Gamma(\gamma)\,\Gamma(\beta-\alpha)}{\Gamma(\gamma-\alpha)\,\Gamma(\beta)}\,,
\\
c_{2}&=&\frac{\Gamma(\gamma)\,\Gamma(\alpha-\beta)}{\Gamma(\gamma-\beta)\,\Gamma(\alpha)}\,.
\end{eqnarray*}
Hence, in our case the leading term of the Hypergeometric function is
\[c_{1}X^{-\alpha}=\frac{\Gamma(2)\,\Gamma(\frac{2\,l}{d-3}+1)}{\Gamma(1+\frac{l}{d-3})\,\Gamma(2+\frac{l}{d-3})}\,X^{\frac{l}{d-3}-1}\,.\]
Substituting into~(\ref{thetil}) we obtain the leading term in $E_{l}$
as $X \to \infty$
\begin{equation} \label{leading}
E_{l} \sim \frac{\Gamma(\frac{2\,l}{d-3}+1)}{\Gamma(1+\frac{l}{d-3})\,\Gamma(2+\frac{l}{d-3})}\,(l-1)\left(\frac{\rho}{\rho_{0}}\right)^{l}.
\end{equation}
Note that the exponent $\sim\rho^{l}$ of the leading term is exactly
what we anticipate for multipole function of order $l$.  The
determination of the prefactor allows one to obtain the constants
$D_{l}$ by matching with the asymptotic zone.

\subsection{The eccentricity}

\FIGURE[t]{\centerline{\epsfig{file=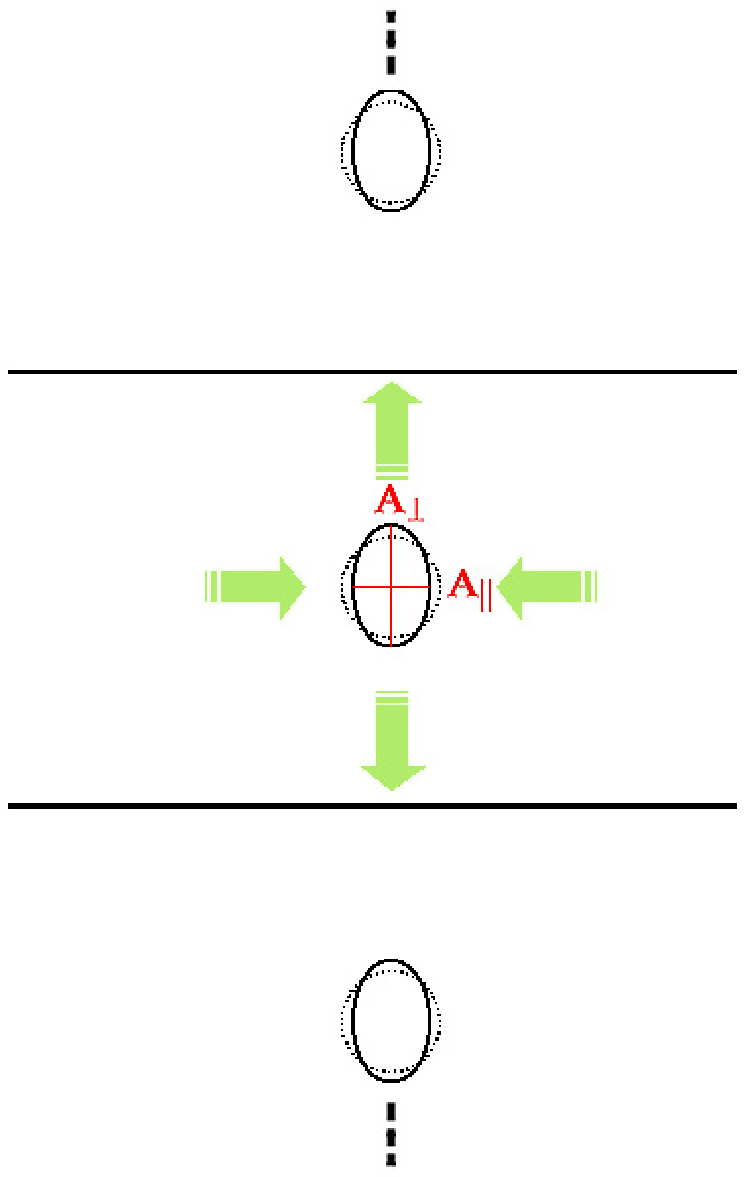, width=.5\textwidth, clip=}}%
\caption{The effective ``mirror" images induce tidal forces that
  stretch the horizon of the black hole. We quantify this by the
  eccentricity which is defined using two sections of the
  horizon.\label{eccentricity}}}

The lowest order deviation from spherical symmetry of the near zone
metric occurs in $n=(d-3)+2=d-1$, where we match the quadrupole (see
figures~\ref{5dialogue} and~\ref{ddialogue}). We can quantify
deviation of the black-hole from spherical symmetry by the
eccentricity of the horizon. We define the eccentricity of the horizon
as in~\cite{KPS1}
\begin{equation} \label{eccen}
\epsilon\equiv\frac{A_{\perp}}{A_{\parallel}}-1\,,
\end{equation}
where $A_{\parallel}$ is the $(d-3)$-dimensional area of the
equatorial sphere at $\chi=\frac{\pi}{2}$ and $A_{\perp}$ is the
$(d-3)$-dimensional area of the polar sphere at $\chi=0,\pi$ (see
figure~\ref{eccentricity}) . Since $g_{\chi\chi}$ is a function only
of $\rho$ and $\chi$ (recall the symmetry $\SO(d-3)$) we obtain
\begin{eqnarray*}
A_{\parallel}&=&\Omega_{d-3}\, g_{\chi\chi}^{\frac{d-3}{2}}
\left(\rho=\rho_{H},\chi=\frac{\pi}{2}\right),
\\
A_{\perp}&=&\Omega_{d-4}\int_{0}^{\pi}g_{\chi\chi}^{\frac{d-3}{2}}(\rho=\rho_{H})\sin^{d-4}(\chi)d\chi\,.
\end{eqnarray*}
Matching the quadrupole in the near zone in order $n=d-1$ to the
newtonian approximation $n=d-3$ (equating the quadrupole terms ---
$l=2$ --- in~(\ref{correction}) and~(\ref{mono2}) using the leading
term~(\ref{leading}) of the quadrupole radial function in the near
zone) we obtain
\begin{equation} \fbox{$\room \displaystyle
D_{2}=\frac{(d-2)\,\zeta(d-1)\,\Gamma(1+\frac{2}{d-3})\,\Gamma(2+\frac{2}{d-3})}{\Gamma(\frac{4}{d-3}+1)}\cdot
\frac{\rho_{0}^{d-1}}{L^{d-1}}\,. $}
\end{equation}
Therefore
\begin{equation} \label{qudroco}
\frac{g_{\chi\chi}^{\mbox{(near)},d-1}}{\rho^{2}}=D_{2}\,
E_{2}(X)\,\Pi_{0}^{2,d}(\chi)\,.
\end{equation}
For the calculation of the eccentricity we need the value of
$E_{2}(X)$ at the horizon, namely, $X=1$. From~(\ref{thetil}) we can
see that
\begin{equation} \label{valueofE}
E_{2}(X=1)=d-3\,.
\end{equation}
As the first non-spherical term is at order $n=d-1$ in the near zone,
the leading term in the eccentricity is proportional to
$\frac{\rho_{0}^{d-1}}{L^{d-1}}$. Thus, we have to substitute
$g_{\chi\chi}^{\mbox{(near)},d-1}$ in the expression for the
eccentricity~(\ref{eccen}) and find the coefficient of
$\frac{\rho_{0}^{d-1}}{L^{d-1}}$ in the expansion. First we expand
\begin{eqnarray}
\left.g_{\chi\chi}^{\frac{d-3}{2}}\,\right|_{\rho=\rho_{H}}&=&\rho_{0}^{d-3}\,\left(1+\cdots +\rho_{0}^{2}\,D_{2}\,(d-3)\,\Pi^{2,d}_{0}(\chi)+\mathcal{O}\left(\frac{1}{L^{d}}\right)\right)^{\frac{d-3}{2}}
 \nonumber\\
&=&\rho_{0}^{d-3}\,\left(1+\cdots +\frac{(d-3)^{2}}{2}\,\rho_{0}^{2}\,D_{2}\,\Pi^{2,d}_{0}(\chi)+\mathcal{O}\left(\frac{1}{L^{d}}\right)\right),
\end{eqnarray}
and at $\chi={\pi}/{2}$ we get
\begin{equation}
\left.g_{\chi\chi}^{\frac{d-3}{2}}\,\right|_{\rho=\rho_{H},\chi=\frac{\pi}{2}}=\rho_{0}^{d-3}\,\left(1+\cdots
-\frac{(d-3)^{2}}{2\,(d-2)}\,\rho_{0}^{2}\,D_{2}+\mathcal{O}\left(\frac{1}{L^{d}}\right)\right),
\end{equation}
where the ellipsis stand for monopole corrections which, of course, do
not contribute to the eccentricity. Substituting in~(\ref{eccen}) we
obtain
\begin{equation}
\epsilon=\frac{(d-3)^{2}}{2}\,\rho_{0}^{2}\,D_{2}\,\left(\frac{\Omega_{d-4}}{\Omega_{d-3}}\int_{0}^{\pi}\Pi^{2,d}_{0}(\chi)\,\sin^{d-4}(\chi)\,d\chi+\frac{1}{d-2}\right)+\mathcal{O}\left(\frac{\rho_{0}^{d}}{L^{d}}\right).
\end{equation}
The integral gives us
\begin{equation}
\int_{0}^{\pi}\Pi^{2,d}_{0}(\chi)\,\sin^{d-4}(\chi)\,d\chi=\frac{1}{d-2}\,\Gamma\left(\frac{d-3}{2}\right)\,\left(\frac{2^{d-2}(d-1)\,\Gamma(\frac{d+1}{2})}{\Gamma(d-1)}-\frac{d\,\sqrt{\pi}
}{\Gamma(\frac{d}{2}-1)}\right).
\end{equation}
This using identities of the Gamma function (see~\cite{erdel1}) yields
the final formula for the eccentricity
\begin{equation} \fbox{$\room \displaystyle
\epsilon=\frac{(d-3)^{4}\,(\Gamma(2+\frac{2}{d-3}))^{2}\,\zeta(d-1)}
{8\,(d-2)\,\Gamma(\frac{4}{d-3})}\cdot\frac{\rho_{0}^{d-1}}{L^{d-1}}
+\mathcal{O}(\frac{\rho_{0}^{d}}{L^{d}}) \,.$}
\label{eccentricity-rho0}
\end{equation}

We can also write the eccentricity as a function of a differently
normalized variable: the relative size of the horizon in conformal
coordinates which we define as (see~\cite{KPS1})
\begin{equation}
\eta \equiv {2\,  \rho_h \over L} = {2^{d-5 \over d-3}\, \rho_0 \over L}\,,
\label{defx}
\end{equation}
where $\rho_h$ is the location of the horizon in conformal coordinates
$ds^2=e^{2\, \hat{A}}\, dt^2 +e^{2\, \hat{B}}\,(dr^2+dz^2) + e^{2\,
  \hat{C}}\, d\Omega_{d-3}^{~2}$ which were used extensively in the
literature and will be used here only for setting the normalization of
$\eta$. $\eta$ is the same quantity denoted by $x$
in~\cite{KPS1,KPS2}. One gets
\begin{eqnarray}
 \epsilon &=& \frac{(d-3)^{4}\,(\Gamma(2+\frac{2}{d-3}))^{2}\,\zeta(d-1)}
{(d-2)\,\Gamma(\frac{4}{d-3})}\cdot 2^{-\frac{(d-4)\,(d+1)}{d-3}}\cdot
\eta^{d-1}+\mathcal{O}(\eta^{d})
\nonumber\\
 & \equiv& \epsilon_1\, \eta^{d-1}\,.
\label{eccentricity-expression}
\end{eqnarray}
The constants $\epsilon_{1}(d)$ are illustrated in
figure~\ref{epsilon1} and in the next table.
$$
\begin{tabular}{|c|c|}
\hline
\emph{d}  & $\epsilon_{1}$ \\
\hline 5 & $\frac{4\pi^{4}}{135}\simeq 2.886$ \\
6 & $\simeq 2.099$ \\
7 & $\frac{\pi^{7}}{2100} \simeq 1.438$\\
8 & $\simeq  0.947$ \\
9 & $\simeq 0.603$ \\
10 & $\simeq 0.375$\\
\hline
\end{tabular}
$$

\FIGURE[t]{\epsfig{file=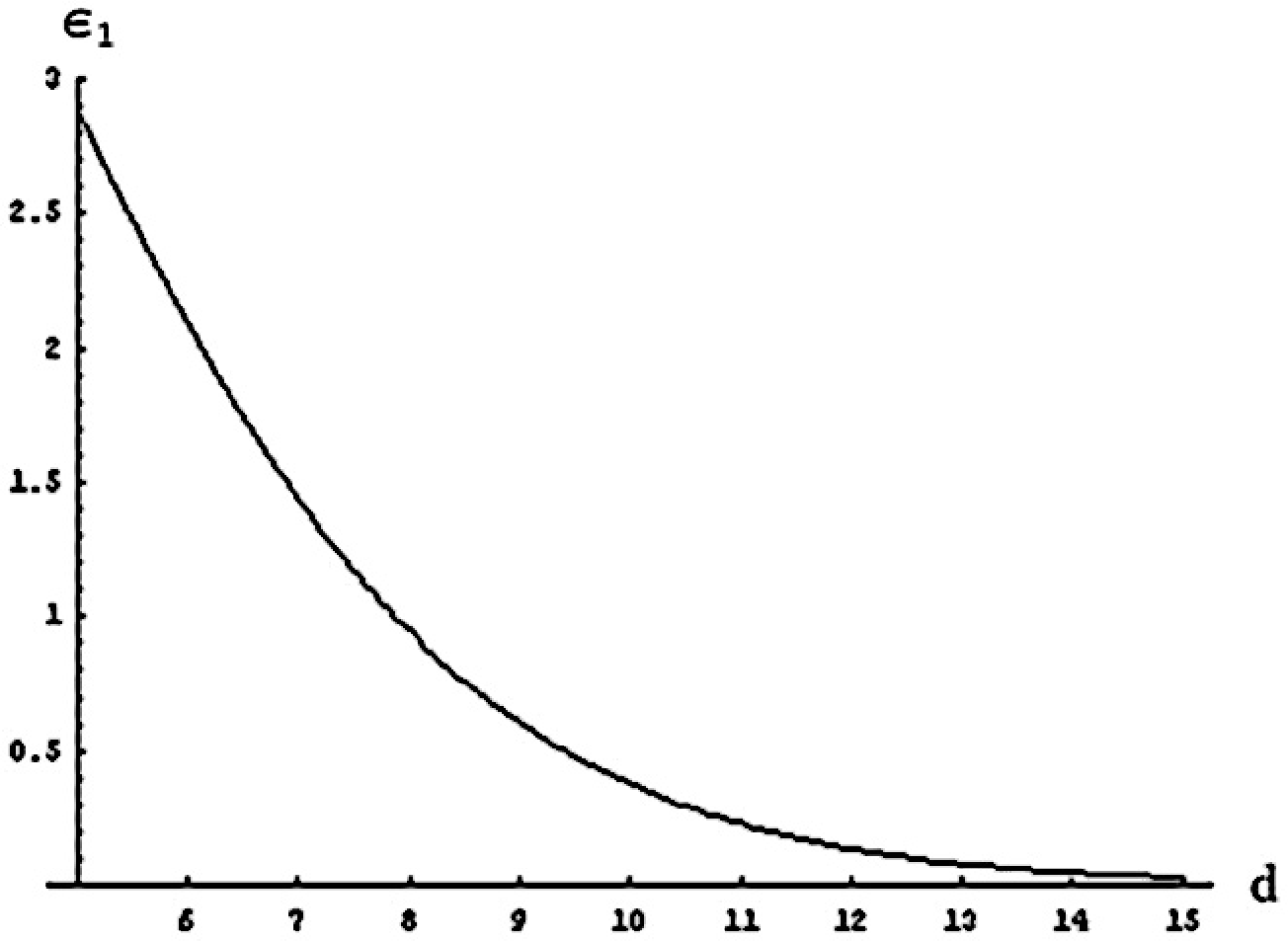, width=.6\textwidth, clip=}%
\caption{A plot of the function $\epsilon_1(d)$ which gives the
  $d$-dependence of the first coefficient of proportionality in the
  expansion of the eccentricity, normalized as in (5.20). Note that
  $\epsilon_1(d)$ is monotonously decreasing to 0.\label{epsilon1}}}

\subsection{The ``Archimedes'' effect}

We define the inter-polar distance $L_{\rm poles}$ to be the proper
distance between the ``poles'' of the black hole measured around the
compact dimension
\begin{equation}
 \Lp = 2\,  \int_{z_H}^{L/2} dz\, \sqrt{g_{zz}}\,,
\end{equation}
where $z_H$ denotes the location of the horizon.  It is convenient to
define a dimensionless parameter out of $\Lp$
\begin{equation}
y \equiv 1-{\Lp \over L} \,.
\label{defy}
\end{equation}
Thus $y$ is the relative decrease in the size of the compact dimension
at $r=0$ due to the presence of the black hole. Like any other
property of the system it is a function of the single dimensionless
quantity which characterizes it, which we choose to be $\eta$, the
relative size of the horizon in conformal coordinates defined
in~(\ref{defx}).

$y(\eta)$ is an interesting quantity to compute in our framework since
it involves both zones in an essential way. The way to compute it is
to pick some mid-point $Z$, divide the integration between the two
zones
\begin{equation}
\half\, \Lp =\int_{z_H}^{Z}dz\,
  \sqrt{g_{zz}^{\mbox{(near)}}} + \int_{Z}^{L/2}dz\,
  \sqrt{g_{zz}^{\mbox{(asymp)}}}\,,
\label{sum-of-patches}
\end{equation}
and confirm that the result is independent of the choice of mid-point
(up to the specified order in $\eta$).

Clearly $y(\eta=0)=0$. Here we wish to compute $y$ to first order in
$\eta$. Which order is required on each zone for the evaluation? Near
the horizon it is enough to take the zeroth order, that is the \Schw\
solution with no corrections (a factor of $\eta$ multiplies it as all
distances scale with $\rho_0$ and the definition of $y$ includes a
division by $L$). In the asymptotic zone one would need in principle
the first order correction (in $\eta$), but since the leading order
newtonian potential is of order $\eta^{d-3}$ and $d \ge 5$ it suffices
to consider again the zeroth order, namely flat compactified space.

Having the matching with the asymptotic zone in mind we write the
\Schw\ metric in the following conformal coordinates
\begin{equation}
 {1 \over \rho_h^{~2}}\, ds^{2~\mbox{(near)}} = -
\( {1 - \psi \over 1 + \psi} \)^2\, dt^2
+ (1 + \psi)^{4 \over d-3}\, \( d\rho^2 + \rho^2\, d
\Omega^2_{d-2} \),
\end{equation}
where
\begin{equation}
\psi = \( \rho_h \over \rho \)^{d-3}.
\end{equation}
These coordinates approach the newtonian gauge as $\rho \gg \rho_h$.

We may now compute the contribution from the near zone
to~(\ref{sum-of-patches}) \begin{equation} {1 \over L}\,
  \int_{z_H}^{Z}dz\, \sqrt{g_{zz}^{\mbox{(near)}}} = \end{equation}
transforming to a dimensionless $\zeta\equiv z/\rho_h$
\begin{equation} = {\rho_h \over L}\, \int_1^{Z/\rho_h} \(1+{1 \over
    \zeta^{d-3}}\)^{2 \over d-3} \, d\zeta = \end{equation} adding
$(+1-1)$ to the integrand in order to separate the part which diverges
with $Z$ \begin{equation} = {\rho_h \over L}\, \left[ \( {Z \over
      \rho_h}-1 \) + \int_1^{Z/\rho_h} \( \(1 + {1 \over \zeta^{d-3}}
    \)^{2 \over d-3} -1 \) d\zeta \right] = \end{equation}
transforming to $w\equiv1/\zeta$ in order to facilitate taking the
limit $\rho_h \sim \eta \to 0$
\begin{eqnarray}
 &=& {Z \over L} + {\rho_h \over L}\, \left[ \int_{\rho_h/Z}^1 \(
 \(1+w^{d-3}\)^{{2 \over d-3}} -1 \)\, {dw \over w^2} -1 \right]
\nonumber\\
 &\simeq& {Z \over L} - {\rho_h \over L}\, I_d \,,
\label{Ihor}
\end{eqnarray}
where the definite (and finite) integral $I_d$ is
\begin{equation}
\fbox{$\room \displaystyle I_d\equiv 1- \int_0^1 \( \(1+w^{d-3}\)^{{2
      \over d-3}} -1 \)\, {dw \over w^2} = 4^{1/k}\, \sqrt{\pi}\,
  {\Gamma\({k-1 \over k}\) \over \Gamma\({1 \over 2} - {1 \over
      k}\)}\,, \qquad k=d-3 ~$}
\label{defId}
\end{equation}
and at this order we neglect the remainder of the integral
\begin{equation}
-{\rho_h \over L}\,  \int_0^{\rho_h/Z} \( \(1+w^{d-3}\)^{{2 \over d-3}} -1 \)\,
 {dw \over w^2}\,.
\end{equation}

The contribution of the (flat metric) asymptotic zone
to~(\ref{sum-of-patches}) is simply
\begin{equation}
{1 \over L}\, \int_{Z}^{L/2}dz\, \sqrt{g_{zz}^{\mbox{(asymp)}}} = {1 \over 2} -
{Z \over L}\,,
\label{Iasymp}
\end{equation}
where due to the low order of the calculation and the choice of
coordinates we can use the same value for $Z$ in both patches, with no
need for corrections arising from a further matching of the patches.

\FIGURE[t]{\epsfig{file=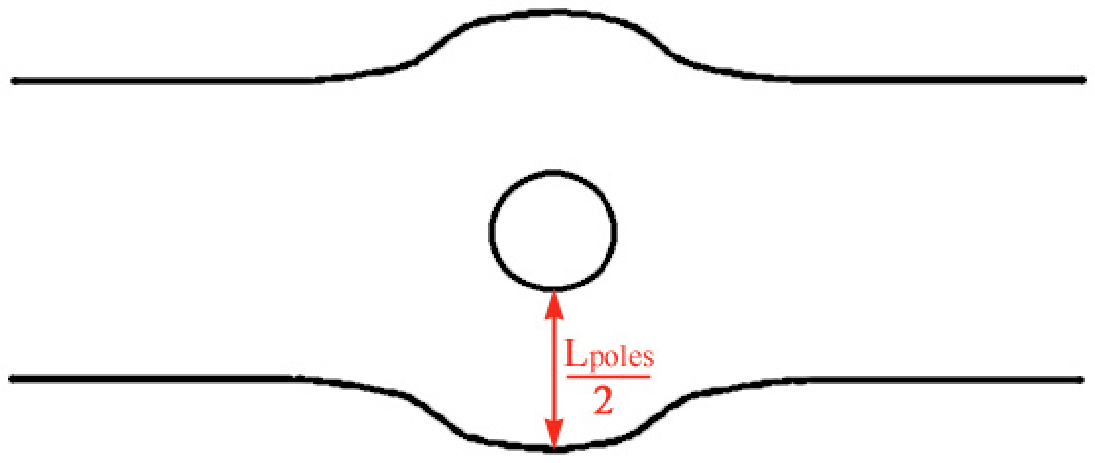, width=.6\textwidth, clip=}%
\caption{Illustration of the ``Archimedes effect". The small black
  hole repels an amount of space proportional to its size. In 5
  dimensions, the black hole repels (to first order) an amount of
  space which is equal to its size (measured in conformal
  coordinates).\label{lpoles}}}

Summing~(\ref{Ihor}),~(\ref{Iasymp}) according
to~(\ref{sum-of-patches}) and using the
definitions~(\ref{defy}),~(\ref{defx}) we confirm that the mid-point
dependence, ($Z/L$), drops (up to the relevant order) and we get our
result
\begin{equation}
 y=2\, I_d\, \eta + o(\eta) \,,
\label{y1}
\end{equation}
where the function $I_d$ is shown in figure~\ref{IdFigure}. It is
monotonously increasing starting from $I_5=0,\, I_6=.6845$, tending to
$1$ as $d \to \infty$. The special behavior at 5d where the ``amount
of space'' outside the black hole remains fixed to first order in the
black hole size was termed ``the black hole Archimedes
effect''~\cite{KPS1} since the black hole seems to ``repel'' as much
space as its own size (figure~\ref{lpoles}).

\FIGURE[t]{\epsfig{file=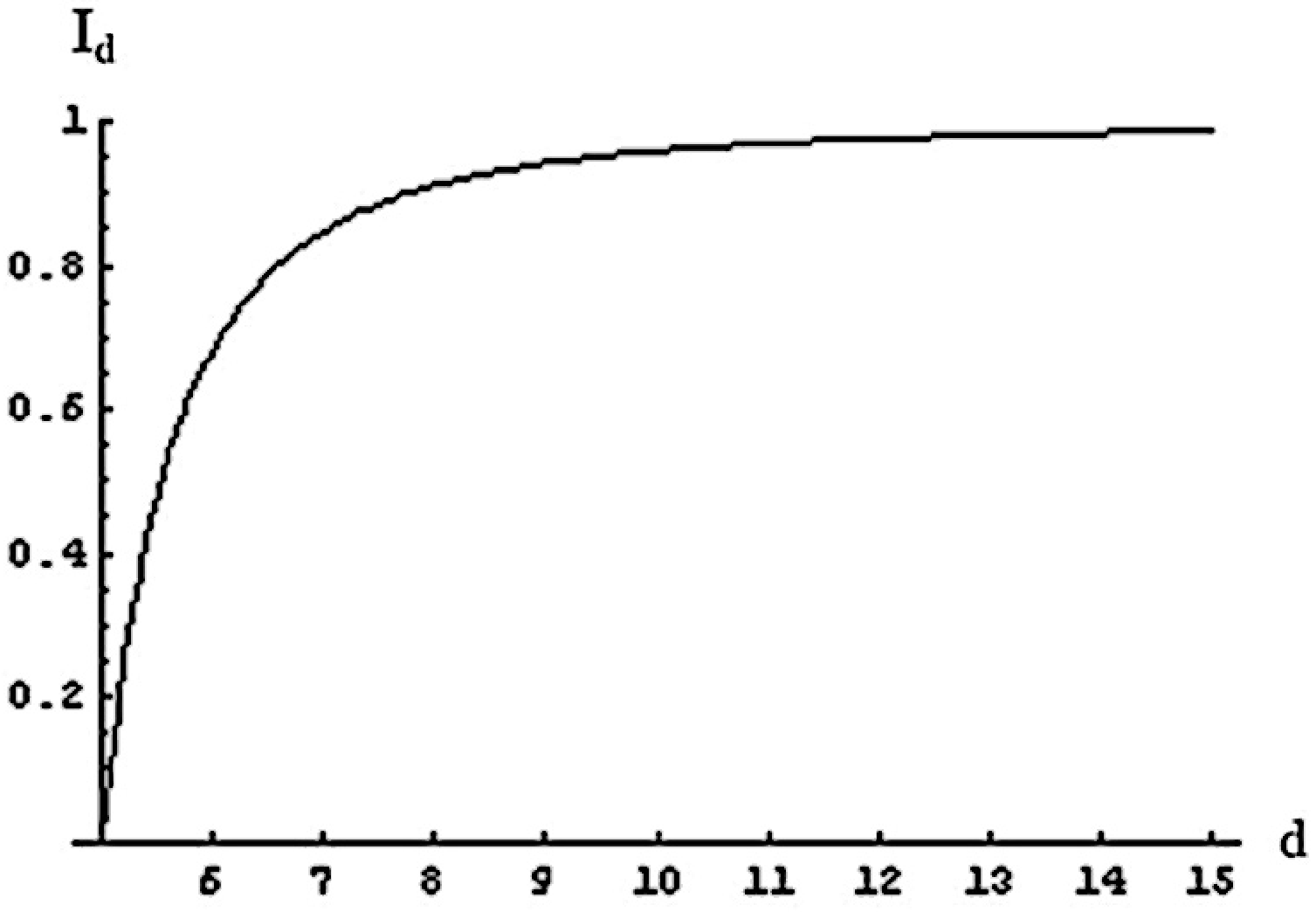, width=.6\textwidth, clip=}%
\caption{The function $I_{d}$ (5.30,5.33) representing the decrease in
  the inter-polar distance (to first order) relative to the size of
  the small black hole (in conformal coordinates). Note that $I_5=0$
  and it rises monotonously to 1.\label{IdFigure}}}

In principle we could have used the corrections to the metric computed
in section~\ref{newtoniansec} and subsection~\ref{MatchingMultipoles}
to improve on this computation and determine the next to leading
correction as well.

\acknowledgments

We would like to thank G. Horowitz and D. Kazhdan each for a
discussion, and E. Sorkin for collaboration on a related problem and
for sharing numerical results. BK thanks Cambridge University,
Amsterdam University and the Perimeter Institute for hospitality
during the course of this work. DG thanks J. Feinberg for a
discussion. This work is supported in part by The Israel Science
Foundation (grant no 228/02) and by the Binational Science Foundation
BSF-2002160.

\appendix
\section{Hypergeometric and Heun's Equations}
\label{ap A} Let us consider a linear ordinary differential
equation of second order
\begin{equation} \label{apSec}
\alpha(x)\,\frac{d^{2}\,f}{dx^{2}}+\beta(x)\,\frac{df}{dx}+\gamma(x)\,f=0\,,
\end{equation}
where $\alpha(x),\beta(x),\gamma(x)$ are holomorphic functions.  When
$\alpha(x_{0})=0$ and $\beta(x_{0})\neq 0$ (or $\gamma(x_{0})\neq 0$)
then $x=x_{0}$ is a singular point of~(\ref{apSec}). This point is
called regular singular point if the limits
\[
\lim_{x\rightarrow x_{0}}
\frac{(x-x_{0})\,\beta(x)}{\alpha(x)}=K_{1}\,,
\] and
\[ \lim_{x\rightarrow x_{0}} \frac{(x-x_{0})^{2}\, \gamma(x)}{\alpha(x)}=K_{2}\,, \]
exist. The regular singular point at infinity is defined in the same
way for $t=0$ in the equation
\[
\frac{d^{2}\,f}{dt^{2}}+\left[\frac{2}{t}-\frac{1}{t^{2}}\,\frac{\beta(\frac{1}{t})}{\alpha(\frac{1}{t})}\right]\frac{df}{dt}+\frac{1}{t^{4}}\,\frac{\gamma(\frac{1}{t})}{\alpha(\frac{1}{t})}f=0\,,
\]
which is obtained after changing the variable $x$ into
$t=\frac{1}{x}$. Equation~(\ref{apSec}) is called a Fuchsian equation
if all the singular points of the equation are regular.

The roots of the equation,
\[\mu^{2}+(K_{1}-1)\,\mu+K_{2}=0\,,\]
are called the \emph{characteristic exponents} at $x_{0}$. There are
two linearly independent solutions to the equation. Thus, at any
singular regular point, there are two different expansions into power
series. If the difference of the exponents is not an integer, the
power series near a singular point $x_{0}$ are of the form
\[(x-x_{0})^{\mu}\sum_{i=0}^{\infty}a_{i}\,(x-x_{0})^{i}\,,\]
where $\mu$ is one of the characteristic exponents. When the
difference of the exponents is an integer one of the expansions may
involve a logarithmic term.

Any Fuchsian equation with 3 singular points can be transformed, by
transformations of the dependent and independent variables, to a
standard form where the singularities are at $0,1$ and $\infty$
\begin{equation}
x\,(1-x)\,\frac{d^{2}\,f}{dx^{2}}+[\gamma-(\alpha+\beta+1)\,x]
\frac{df}{dx}-\alpha\, \beta \,f=0\,,
\end{equation}
where $\alpha,\beta,\gamma$ are parameters.  The equation in this form
is called the Hypergeometric equation. The solutions of~(\ref{apSec})
can be characterized completely by the singularities and the
corresponding characteristic exponents at each singularity. There
exists a compact scheme to summarize the information about the
solutions of the Hypergeometric equation --- the Riemann
\emph{P}-Symbol (see for instance~\cite{erdel1,spe1})
\[
P \left (\begin{array}{ccccccc}
0 \;& & 1 \; & & \infty \;& &\\
0 \;& & 0 \;& & \alpha \;& & ;\ x \\
1-\gamma \; & & \gamma-\alpha-\beta \;& & \beta \;& &
\end{array}
\right ).\]
The first row of the matrix indicates the three regular singular
points. The two numbers beneath each singular point are the
characteristic exponents at each singular point. The expansion around
$0$ of the solution with a zero exponent is the Hypergeometric
function
\[_{2}\!F_{1}(\alpha,\beta,\gamma;x)=\sum_{n=0}^{\infty}\frac{(\alpha)_{n}\,(\beta)_{n}}{(\gamma)_{n}}\,x^{n}\,,\]
where $(\alpha)_{n} \equiv \frac{\Gamma(\alpha+n)}{\Gamma(\alpha)}$
when $\alpha$ is different from a negative integer. The other 5
expansions can be obtained by transformations of this Hypergeometric
function.

A Fuchsian equation with 4 singular points can be transformed
similarly to a canonical form where the singularities are at $0,1,a$
and $\infty$ ($a$ can be any point in the complex plane different from
the other singular points). In this canonical form the equation is
called Heun's equation~\cite{Heun,erdel}
\begin{equation}
\frac{d^{2}\,f}{dx^{2}}+\left(\frac{\gamma}{x}+\frac{\delta}{x-1}+\frac{1+\alpha+\beta-\gamma-\delta}{x-a}\right)\,\frac{d\,f}{dx}+\frac{\alpha\,\beta
\,x -q}{x\,(x-1)\,(x-a)}\,f=0\,.
\end{equation}
This type of equation is a generalization of the Hypergeometric
equation. Then in a similar manner a Riemann \emph{P}-Symbol can be
constructed for the 4 singular points:
\[
P \left (\begin{array}{ccccccccc}
0 \;& & 1 \;& & a \;& & \infty \;& &\\
0 \;& & 0 \;& & 0 \;& & \alpha \;& & ;\ x,\;\;\; q\\
1-\gamma \;& & 1-\delta \;& & \gamma+\delta-\alpha-\beta \;& &
\beta \;& &
\end{array}
\right ).\]

Unlike the Hypergeometric equation, Heun's equation is not
characterized completely by its characteristic exponents at each
singular point. There is another parameter --- $q$. This parameter is
called the \emph{accessory} or \emph{auxiliary} parameter.

\section{The surface gravity}
\label{ap B}

Let us compute the surface gravity for a metric of the form
\[ds^{2}=-f(\rho)\,dt^{2}+g(\rho)d\rho^{2}+\sigma_{ij}\,dx^{i}\,dx^{j}\,,\]
(where $\sigma_{ij}\,dx^{i}\,dx^{j}$ stands for the rest of the
spatial part of the metric). Using the definition of surface
gravity~\cite{wald2}
\[\kappa^{2}=-\frac{1}{2}(D^{\mu}\zeta^{\nu})\,(D_{\mu}\zeta_{\nu})\,,\]
where $\zeta^{\nu}$ is the Killing vector field $\del_t$. Using
the fact that
\begin{eqnarray*}
\Gamma_{t \rho}^{t}&=&\frac{f'}{2\,f}\,,
\\
\Gamma_{t t}^{\rho}=\frac{f'}{2\,g}\,,
\end{eqnarray*}
we get the following formula for the surface gravity
\begin{equation} \label{kap}
\kappa=\frac{1}{2} \left.\frac{ \mid f'(\rho) \mid
}{\sqrt{f(\rho)\,g(\rho)}}\,\right|_{\rho=\rho_{H}} .
\end{equation}

\section{5d scalar harmonics}
\label{ap C}

For concreteness, let us write down a basis for vector harmonics in 5
dimensions. First, we have a natural basis element for vector
harmonics which is derived from the scalar spherical harmonics
\[
U_{\mu}\,Q^{n}_{lm} \equiv \hat{D}_{\mu} Q^{n}_{lm} \,,
\]
namely $\hat{D_{\mu}}$ is the covariant derivative on $S^{3}$
\[
\hat{D}_{\mu} \equiv (\hat{D}_{\chi},\hat{D}_{\theta},\hat{D}_{\varphi}) \,.
\]
This basis element is a vector, whose direction is defined by the
gradient on the sphere. We look for another two vector basis elements
on $S^{3}$ which are orthogonal to this one.\footnote{The
  orthogonality is defined using the natural inner product; for two
  vector operators $A_{\mu}$ and $B_{\nu}$ we define $<a,B> \equiv
  \int\! \int \!\int \!(A_{\mu} Q^{n}_{lm})\, \gamma^{\mu\nu}\,
  (B_{\nu} Q^{n}_{lm})\,\sqrt{\gamma}\, d\theta \,d\varphi\, d\chi$
  where $\gamma_{\mu\nu}$ is the natural metric induced on the
  sphere.} We fix the arbitrariness in the choice of the other two
vectors by considering the $\SO(3)$ symmetry in the $(\theta,\varphi)$
coordinates. Namely, we choose a second vector basis element which is
orthogonal to the first one in the ($\theta,\varphi$) plane
\[V_{\mu}\,Q^{n}_{lm}\,,\]
where
\[V_{\mu}\equiv\sin(\chi)\,\epsilon_{\mu}^{\nu}\,U_{\nu}=
(0,\frac{\sin\chi}{\sin\theta}\,\hat{D_{\varphi}},-\sin\theta\,\sin\chi\,\hat{D_{\theta}})\,.\]
Here $\epsilon_{\mu\nu}$ represent the two dimensional Levi-Civita
tensor for the coordinates $\theta-\varphi$ and zero for components
with $\chi$ coordinate. The multiplication by $\sin(\chi)$ is due to
the difference between the inner products on $S^{2}$ and $S^{3}$.

The third basis element is defined as a vector which is orthogonal to
the other two basis elements
\[W_{\mu}\,Q^{n}_{lm}\,,\]
where
\[W_{\mu}\equiv\epsilon_{\mu}^{\nu\rho}\,U_{\nu}\,V_{\rho}=(\frac{l\,(l+1)}{\sin(\chi)},\hat{D_{\chi}}(\sin\chi\,\hat{D_{\theta}}),\hat{D_{\chi}}(\sin\chi\,\hat{D_{\varphi}}))\,,\]
and $\epsilon_{\mu\nu\rho}$ is the three-dimensional Levi-Civita
tensor. (In the calculation we used the eigenvalue equation for the
spherical harmonics on $S^{2}$.) Gerlach and Sengupta~\cite{gerlach}
obtained exactly this form of a basis of vector harmonics on $S^{3}$
using a different method.

\end{document}